\documentclass{aa}
\usepackage{graphics,psfig,latexsym}
\usepackage{times,graphicx,epsfig}
\newcommand{\bq}{\begin{equation}}
\newcommand{\eq}{\end{equation}}
\newcommand{\bqn}{\begin{eqnarray}}
\newcommand{\eqn}{\end{eqnarray}}
\newcommand{\dd}{\mbox{\rm d}}
\newcommand{\msun}{\rm{M}_\mathrm{\rm \sun}}

\begin{document}
\title{An evolutionary disc model of the edge-on galaxy NGC 5907}

\author{Andreas Just\inst{1}\and
        Claus M\"ollenhoff \inst{2}\and
        Andrea Borch \inst{1}}

\offprints{A. Just}
\mail{just@ari.uni-heidelberg.de}
\institute{Astronomisches Rechen-Institut am ZAH, M\"onchhofstra{\ss}e 12-14, 
69120 Heidelberg, Germany
\and
Landessternwarte am ZAH, K\"onigsstuhl 12, 69117 Heidelberg, Germany}

\date{Printed: \today}

 
  \abstract
{We present an evolutionary disc model of the edge-on galaxy NGC 5907 based on a
continuous  star formation history and a continuous dynamical heating of
the stellar subpopulations.} 
{This model explains the disparate two
observational facts: 1) the exponential vertical disc 
structure in the optical and NIR of the non-obscured part of the stellar disc 
and 2) the FIR/submm luminosity enhanced by about a factor of four near the 
obscured mid-plane, which requires additional dust and also stellar light 
to heat the dust component.} 
{We use multi-band photometry in U, B, V, R, and I- band combined with radiative 
transfer through a dust component to simultaneously fit the vertical 
surface-brightness and
colour index profiles in all bands adopting a reasonable star formation history and
dynamical heating function. The vertical distribution of the stellar
subpopulations are calculated self-consistently in dynamical equilibrium and
the intrinsic stellar emissivity is calculated by
stellar population synthesis.} 
{The final disc model reproduces the surface-brightness profiles in all bands with a
moderately declining star formation rate and a slowly starting heating function
for young stars. The total dust mass is $5.7\times 10^7\,\msun $ as 
required from the FIR/submm measurements. Without a recent star burst 
we find in the midplane an excess of 5.2-, 4.0-, and 3.0-times 
more stellar light in the U-, B-, and V-band, respectively. 
The corresponding stellar mass-to-light ratios are 0.91 in
V- and 1.0 in R-band. The central face-on optical depth in V-band is 
$\tau_\mathrm{V}^{f}=0.81$ and the radial scale length of the dust is 40\% larger than
that of the stellar disc.}
{Evolutionary disc models are a powerful method to understand the vertical
structure of edge-on galaxies. Insights into the star formation history and the
dynamical evolution of stellar discs can be gained.
FIR/submm observations are necessary to restrict
the parameter space for the models.}

\keywords{Stellar dynamics -- Galaxies: kinematics and dynamics -- 
Galaxies: interactions -- dark matter}

\maketitle



\section{Introduction\label{intro}}

\subsection{Vertical structure of stellar discs}\label{vertstruct}

The vertical structure of the stellar disc in late-type galaxies is mainly 
determined
by the star formation history combined with the dynamical heating
$\sigma(t)$ of the stellar population due to gravitational
perturbations. Young stellar subpopulations are confined tightly to the
midplane with small velocity dispersion, whereas the older subpopulations are
distributed over larger heights due to their higher velocity dispersion. This
basic feature is well-known in the Galaxy (Wielen \cite{wie77}, Freeman
\cite{fre91}, Edvardsson et al. \cite{edv93} e.g.). Dynamical heating leads
to an increasing mean age of the population with increasing height above the
midplane. This results in an increasing mass-to-light ratio and vertical colour index
gradients from blue to red. In principle, these colour index gradients should be
directly observable in
edge-on galaxies but unfortunately, strong dust extinction and reddening
often dominates the vertical colour index distribution. Therefore detailed models are
necessary to extract the intrinsic stellar disc properties from
multi-colour photometry.

In a series of papers van der Kruit \& Searle (\cite{vdk81,vdk82a,vdk82b})
started to investigate the vertical and radial properties of the surface-brightness
profiles of edge-on galaxies. They determined for a number of nearby galaxies
(including NGC 5907) the radial scale length, the radial cutoff, and the
vertical scale length. They recognised that the shape of the vertical 
surface-brightness profiles can be well fitted by a $sech^2$-model disregarding the
central minimum in the case of strong dust lanes. In recent decades much work
has been done on the phenomenological analysis of edge-on galaxies
 (e.g. Guthrie \cite{gut92}, Pohlen et al. \cite{poh02}). 
 
In our Galaxy the stellar subpopulations of the thin disc are not homogeneously
mixed. The vertical thickness of the  subpopulations increases with age 
(Wielen \& Fuchs \cite{wie88}). Wainscoat et al. (\cite{wai89}) observed a
similar vertical structure in the edge-on galaxy IC 2531. For the analysis
 of the intrinsic physical structure of galactic discs 
we started to investigate the vertical surface-brightness and colour index profiles of edge-on
galaxies in order to analyse the age distribution of stellar discs
(Wielen et al. \cite{wie92}, Just et al. \cite{jus96}). We used evolutionary
stellar population models for the intrinsic light distribution and include dust
extinction in the radiative transfer for the observable vertical surface-brightness
profiles. We found that in edge-on galaxies the vertical colour index gradients above
the innermost part, where extinction dominates, are a signature of 
the changing properties of the stellar population with height above the midplane.
De Grijs and Peletier (\cite{dgr00}) also found from a large sample of
edge-on galaxies that the stellar
population properties in the discs can be seen in the vertical colour index 
gradients above the extinction features.

For NGC 5907 and other nearby edge-on galaxies Xilouris et al.
(\cite{xil97,xil99}) used an extensive radiative transfer code including
scattered light to model the two-dimensional light distribution of the stellar
discs. Using exponential intrinsic luminosity distributions different for each
band (B,V,I in the case of NGC 5907) and an exponential disc for the dust
component, they determined radial and vertical scale lengths of the star light
and the properties of the dust component. The resulting relative low dust mass 
corresponds to a gas-to-dust ratio of 810 (corrected for the Helium contribution). 
In recent years FIR observations offered a new, more
direct way to deduce the dust properties of galaxies. Popescu et al. (\cite{pop00}) 
developed a sophisticated temperature model of the dust component in edge-on
galaxies in order to fit the FIR observations and applied it to NGC 891. 
With this method Misiriotis et  al. (\cite{mis01}) determined for NGC 5907 a dust
mass 4 times larger than Xilouris et al. (\cite{xil99}). For the source of the
FIR excess they adopted an additional young stellar component in a very thin
disc, which is completely hidden by dust extinction in the optical 
and NIR bands.
Recently Stevens et al. (\cite{ste05}) used
a two-temperature dust model to fit the IRAS and SCUBA data for a sample of spiral
galaxies. For NGC 5907 they determine a dust mass only three times larger 
than the Xilouris et al. value.

\subsection{Content of this paper}\label{content}

\begin{figure}[t]
\hbox{\hspace{0cm}              
 \psfig{figure=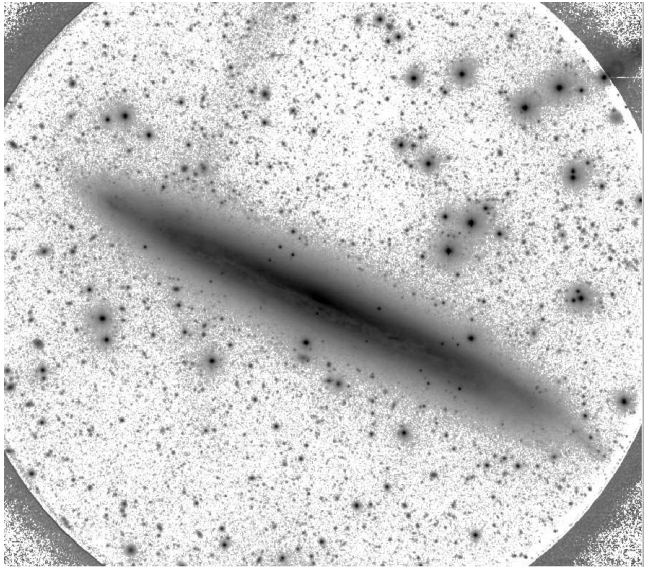,width=8.5cm,clip=}}
\caption[]{Deep R image of NGC 5907, sum of 4 exposures.
The field of view is $15\times 13.25$ arcmin. N is top, E is left.
(cf Sect.~\ref{observ})}
\label{figimagr}
\end{figure}

In this paper we present an evolutionary stellar disc model for NGC 5907, 
which 
explains both observational results: 1) The exponential vertical profiles in all
optical bands  are reproduced by the outer stellar disc dominated by the older
thin disc population and are no longer just a set of independent fit functions in
each band as in Xilouris et al. (\cite{xil99}).
2) The large amount of dust determined from FIR/submm observations by Misiriotis
et al. (\cite{mis01}) exceeding the dust component found in Xilouris et al. by a
factor of four is necessary in our model to obscure the additional light of the
young stellar subpopulations with low mass-to-light ratio near the midplane of the
disc. There is no additional recent star burst required to explain the
additional star light needed for the dust heating.

The basis of our disc model is a vertical cut with a self-consistent detailed
distribution of stellar subpopulations. The vertical distribution of the stars
is determined by the dynamical equilibrium of the stellar subpopulations
described by the star formation history of the disc 
and by the dynamical heating function 
(measured by the increasing velocity dispersion).
The gravitational forces of the gas component and the dark matter halo are
included. 
The intrinsic stellar light distribution is calculated from population 
synthesis models of simple stellar populations (SSP). The resulting
 vertical profile is
then extended to an exponential disc up to the cutoff radius $R_\mathrm{max}$ by
scaling the surface densities with constant scale heights. The intrinsic
structure is not changed as a function of radius. We compute the observable
vertical surface-brightness and colour index profiles by radiative transfer calculations with
an exponential dust component, where the inclination of the disc 
with respect to the line of sight is fitted.

The parameters of the basic vertical profile are chosen to match all constraints 
at some fixed radius $R_0$ of the disc, where the disc is dominating the light
and the bulge can be neglected. At that radius we fix the relative
contribution of the gas and dark matter component 
$Q_\mathrm{g}=\Sigma_\mathrm{g}/\Sigma_\mathrm{tot}$ and 
$Q_\mathrm{h}=\Sigma_\mathrm{h}/\Sigma_\mathrm{tot}$ to the total
surface density $\Sigma_\mathrm{tot}$ up to a  maximum height 
$z_\mathrm{max}$ above the
midplane. We will use
$R_0=10\,\mathrm{kpc}\approx2\,R_\mathrm{s}$ with radial scale length 
$R_\mathrm{s}$ of the stellar 
disc, which is near the maximum of 
$\Sigma_\mathrm{d}/\Sigma_\mathrm{h}$ in the case of an isothermal halo. 
The scale height of the gas component relative to the effective scale
height of the stars $z_\mathrm{g}/z_\mathrm{0}$ and 
$s_\mathrm{h}=\sigma_\mathrm{h}/\sigma_\mathrm{e}$, the ratio of the velocity 
dispersion of the dark matter halo and the maximum velocity dispersion of the
stars are also determined at $R_0$.

In an iterative process the crucial fitting functions $SFR(t_\mathrm{a}-t)$ and 
 $\sigma(t)$ are optimised together with the global parameters: inclination
 $i$, radial scale length $R_\mathrm{s}$, effective scale height $z_\mathrm{s}$, dust
 distribution, and total stellar and dust mass. For comparison with observations
 we use a series of vertical surface-brightness (and colour index) profiles parallel to the
 minor axes in U,B,V,R,I bands, which we obtained from corresponding deep 
photometric images of
 NGC 5907. The same type of intrinsic disc model was also successfully applied
 to the solar neighbourhood in order to analyse the connection between the local
 velocity distribution 
 functions of the main sequence stars from the Catalogue of Nearby Stars 
 (CNS4, Jahreiss \& Wielen \cite{jah97}) and the
 vertical density profiles of these stars (see Just \cite{jus01,jus02,jus03}).

In Sect. \ref{param1} we collect the basic parameters of NGC 5907, which are
necessary for the basic scaling of our disc model to the observational data.
In Sect. \ref{components} we describe the building blocks of our disc
model. 
Sect. \ref{project} gives the iteration procedure with details on the scaling
method, radiative transfer and the multi-colour analysis.
In Sect. \ref{observ} our observations and the derivation of vertical colour index and
surface-brightness profiles are presented.
Fig.~\ref{figimagr} shows a deep R image (cf Section~\ref{observ} as an example
of for the high quality of data. 
In Sect. \ref{result} we show the final comparison of the profiles, give the
global parameters of our best model and discuss in detail the intrinsic
structure of the stellar disc.
A discussion of some crucial aspects follow in Sect. \ref{discuss}.
Sect. \ref{conclusion} contains the concluding remarks.

\section{Basic parameters of NGC 5907 \label{param1}}

Here we discuss the basic scaling parameters of NGC 5907, namely distance, 
radial and
vertical scale length and inclination of the stellar disc, and the masses of 
the stellar, gaseous and dark matter component.

\subsection{Distance to NGC 5907\label{dist}}

There is some uncertainty in the distance determination of nearby
galaxies from the Hubble expansion, because the peculiar motion of the galaxies
is not negligible. In this work we use a distance of $D=11$\,Mpc with 
$V_\mathrm{GSR}=817$\,km/s from the RC3 catalogue (de Vaucouleurs et al. \cite{dev91})
and a Hubble constant of $H_0=75$\,km/s/Mpc. The resulting scale conversion is
$1\arcsec=53.3$\,pc and the distance modulus is $\Delta m=-30.2$\,mag.

Another distance determination for NGC~5907 using the
Tully-Fisher relation leads to a slightly larger distance of
 $D$=11.6 or 12.0\,Mpc (Sofue \cite{sof96,sof97}). In this method the main
 uncertainty arises from the conversion of the edge-on luminosity to a face-on
 luminosity. Zepf et al. (\cite{zep00}) used a distance of $D=13.5\pm 2.1$\,Mpc
 from a combination of the R-band Tully-Fischer relation with a peculiar motion
 model and discussed also (due to the lack of resolved giant stars in the outer
  bulge region from HST observations) the
 possibility of a significantly larger distance.

\subsection{Disc parameters \label{sdisc}}

For the outer parts of the disc, where extinction is not significant, 
the brightness distribution of the disc of NGC 5907 can be modelled by
 exponential profiles in the vertical and radial direction. The radial scale
 length increases from 3.7\,kpc in H-band (Barneby \& Thronson \cite{bar94}) to
 more than 5\,kpc in the blue 
 (van der Kruit \& Searle \cite{vdk82a}: 5.7\,kpc in J-band,
Xilouris et al. \cite{xil99}: 5.02\,kpc in B-band). 
Van der Kruit \& Searle also determined the cutoff radius of the exponential
disc at $R_\mathrm{max}=19.3$\,kpc, which has some influence on the inner parts of the
radial surface-brightness profiles due to projection effects. 
The cutoff was not included in the model of Xilouris et al. (\cite{xil99}) leading 
to an underestimation of the intrinsic scale length.

The effective scale height of the disc is approximately 
independent of radius, but depends strongly on the model used, i.e. additional
components like thick disc, bulge and also on the radial
profile for edge-on galaxies. From moderately deep photometry
van der Kruit \& Searle found $z_\mathrm{s}=410$\,pc in J/band and
Xilouris et al. $z_\mathrm{s}=340$\,pc in B-band, and Barneby \& Thronson 
$z_\mathrm{s}=320$\,pc in H-band,
which are relatively small values. With a One-disc-model for the very deep 
R-band photometry Morrison et al. (\cite{mor94}) determined a scale height of
$z_\mathrm{s}=467$\,pc.

From the shape of the outer isophotes van der Kruit \& Searle found an 
inclination of $i=87.0\degr $, which is widely used as a standard value. Xilouris et
al. determined independently a value of $i=87.2\degr $. 
Morrison et al. (\cite{mor94}) observed a stellar warp in the outer parts of the
disc. Therefore the intrinsic inclination may be even
higher, since the warp smears out the isophotes
additively.

Since in our disc model we assume only one stellar disc for all five observed bands, 
we redetermine the radial and vertical scale length, the cutoff radius and the
inclination in the fitting procedure.

\subsection{Masses \label{mass}}

For the construction of the disc model we need the relative
surface densities of the different components. The scaling to the absolute
values is then a result of the surface-brightness profile fitting 
(cf Sect. \ref{fitting}), and the total masses of
the stellar and gaseous components follow from the radial extrapolation
to exponential discs. We use the total hydrogen mass $M_\mathrm{H}$ and convert it to
the total gas mass $M_\mathrm{g}$ which will be used for the determination of the 
gravitational force and to calculate the gas-to-dust ratio $F$. 
For the dark matter halo we need only a reliable estimate of
the local density in the disc.

\subsubsection{Gas masses \label{gmass}}

The atomic gas mass is determined from the 21cm luminosity.
We use the standard value
\bq
M_\mathrm{HI}=6.9\times10^9\msun
\eq
 (Dumke et al. \cite{dum97}, Stevens et al. \cite{ste05}). The molecular
gas mass is much more uncertain. It is derived by converting CO measurements
using the conversion factor $X$, which gives the corresponding $H_2$ mass.  
Dumke et al. (\cite{dum97}) determined individually for the galaxy NGC 5907 
a reduced conversion factor $X$ yielding $M_\mathrm{H_2}=0.9\times10^9\msun$. 
The standard $X$-value would have led to $1.8\times10^9\msun$, 
which demonstrates the intrinsic uncertainty of the molecular gas
mass determination.
We use the molecular gas mass redetermined by 
Stevens et al. (\cite{ste05}) of
\bq
M_\mathrm{H_2}=1.7\times10^9\msun
\eq
 (rescaled to the distance of $D=11\,Mpc$). For the contribution of He to the
total gas mass we add 40\% to the total hydrogen mass 
$M_\mathrm{H}=M_\mathrm{HI}+M_\mathrm{H_2}$. This leads to the total gas mass of
\bqn
M_\mathrm{g}&=&1.4\,M_\mathrm{H}=1.2\times10^{10}\msun \quad\mbox{with}\label{eqmgas}\\
M_\mathrm{H}&=&M_\mathrm{HI}+M_\mathrm{H_2}=8.6\times10^{9}\msun \quad .
\eqn

The radial distribution of HI is flatter and more extended than the 
stellar light distribution and is not exponential. The CO distribution is much
more confined to the inner parts than that of HI (Dumke et al. \cite{dum97}). 
For simplicity we will also adopt for the scaling of the gas fraction an 
exponential disc, but allow for a scale length larger than the radial scale 
length of the stellar disc.
We will use the same spatial distribution as for the dust component,
which is determined by our fitting procedure.

\subsubsection{Stellar and dark matter masses \label{smass}}

The determination of the stellar disc and dark matter (DM) halo mass from the 
rotation curve
is very uncertain due to the well-known disc-halo degeneracy. 
Sackett et al. (\cite{sac94}) constructed three-component mass models (HI, stellar
disc, dark matter halo) to fit the
HI-rotation curve of Sancisi and van Albada (\cite{san87}). They found equally
good fits with disc mass-to-light ratios of 1, 2, and 4 in the R-band.
The models of Sackett et al. with $M/L=1$ and $M/L=2$ yield a range of surface
density ratios of stellar disc and halo of
\bq
\left.\frac{\Sigma_\mathrm{s}}{\Sigma_\mathrm{h}}\right|_\mathrm{R_0}=2\dots 5
\quad\mbox{with}\quad
R_0=10\,\mathrm{kpc},\,z_\mathrm{max}=5.1\,\mathrm{kpc}.
\eq
Typical mass-to-light ratios of stellar discs in the V-band are of order unity
and are dependent on the average age of the population. In the solar
neighbourhood we have $M/L_\mathrm{V}=0.78$ and the extrapolation to the solar cylinder
is $M/L_\mathrm{V}\approx 1.4$ for the mass-to-light
ratio of the surface density (from CNS4 data, Jahreiss 
\& Wielen \cite{jah97}). Our final model yields $M/L_\mathrm{V}=0.9$ and $M/L_\mathrm{R}=1.0$.
Therefore we will use the $M/L_\mathrm{R}=1$ model of Sackett et al. (\cite{sac94})
for the scaling of the DM halo.

Sofue (\cite{sof96}) presented the joint rotation curve of CO and HI, which shows
in HI a slightly higher and more pronounced maximum, which is harder to
reconstruct with a low mass disc of $M/L_\mathrm{R}<2$. 
From H-band photometry Barneby and Thronson (\cite{bar94}) investigate mass
models including a flattened bulge and determined a small bulge with scaling
radius $0.23\,kpc$ and total mass of $M_\mathrm{B}=9\times 10^{9}\,\msun $.
The CO rotation curve of Sofue (\cite{sof96}) confirms the kink produced by the
bulge at $V_\mathrm{c}(R=1\,$kpc$)\approx 200$\,km/s of model b) in 
Barneby and Thronson (\cite{bar94}). As a consequence the pollution of the disc
luminosity with bulge light at radial distances larger than 3\,kpc is very
small. Therefore we decided to neglect the bulge component in our
analysis in order to keep the number of free parameters small.

\subsubsection{Dust mass and extinction\label{dust}}

Our aim is to construct a disc model with a dust component, which has a mass
comparable to 
\bq
M_\mathrm{d}=6\times10^7\msun\,,
\eq
the value determined by Misiriotis et al. (\cite{mis01}). 
 We use the standard extinction law of Rieke \&
Lebofsky (\cite{rie85}), which was confirmed by Xilouris et al. (\cite{xil99})
for NGC 5907 and which also goes into the conversion factor of extinction to dust
mass.
In our model we determine the
spatial distribution of extinction $A_\mathrm{V}(R,z)$ (see Eq. \ref{eqdust}). This
will be converted to the dust mass distribution by
\bq
\rho_\mathrm{d}=0.175\times 10^{-3} A_\mathrm{V} \msun\,\mathrm{pc}^{-3} \label{eqrhod}
\eq
with $A_\mathrm{V}$ in [mag/kpc]. This is the same conversion factor as used by
Xilouris et al. (\cite{xil99}) and Misiriotis et al. (\cite{mis01}).

\section{Building blocks of the disc model \label{components}}

The aim of this work is to construct a physical model of the stellar disc
in order to reproduce the vertical surface-brightness and colour index profiles from U,B,V,R,
and I-band observations. 
Therefore we pay much attention to the internal structure of the stellar disc.
The gas and dust component and the dark matter halo are modelled in a simple way.
The bulge contribution is neglected, because the number
of free parameters and fitting functions would be approximately doubled with
very small effect on the parameters of the disc. A thick disc component is
excluded by very deep photometry and the faint stellar halo is below our
observational limit (Morrison et al \cite{mor94}).
 We apply our model to those
regions of the disc where the thin disc strongly dominates.

In this section we describe the different ingredients necessary to construct the
disc model and compute their intrinsic properties. The projection onto the sky and
the fitting procedure is given in Sect. \ref{project}.

\subsection{Self-gravitating disc \label{grav}}

The backbone of the disc is a self-gravitating vertical disc profile including 
the gas component in the thin disc approximation. In this approximation the
Poisson-Equation is one-dimensional
\bq
\frac{\dd^2\Phi_\mathrm{self}}{\dd z^2}=4\pi G \rho(z) \quad,\label{eqpoi}
\eq
where $\rho(z)$ is the self-gravitating density and $\Phi_\mathrm{self}(z)$ is the
corresponding potential. Since we want to construct the disc in dynamical 
equilibrium, the density of the sub-components will be given as a function 
of the total potential $\rho_\mathrm{j}(\Phi)$ and not of height $z$.
In the case of a purely self-gravitating thin disc
(with $\Phi_\mathrm{self}=\Phi$, i.e. no external potential) the Poisson equation can be
integrated leading to
\bq
\left(\frac{\dd\Phi}{\dd z}\right)^2
        =8\pi G \int_0^{\Phi}\rho(\Phi')\dd\Phi' \quad.\label{eqkz}
\eq
Then the vertical distribution is given by the implicit function $z(\Phi)$ 
via direct integration
\bq
z=\int_0^{\Phi}\dd\Phi'
        \left[8\pi G \int_0^{\Phi'}\rho(\Phi'')\dd\Phi''\right]^{-1/2}
         \quad.\label{eqz}
\eq
If an external potential is included, an iteration process  is  necessary to
solve for the vertical distribution. In order to avoid the iteration at that
point, we model all gravitational components by a thin disc approximation.
We include in the total
potential $\Phi$ the stellar component $\Phi_\mathrm{s}$, 
the gas component $\Phi_\mathrm{g}$ and the dark matter halo contribution $\Phi_\mathrm{h}$
\bq
\Phi(z)=\Phi_\mathrm{s}(z)+\Phi_\mathrm{g}(z)+\Phi_\mathrm{h}(z)\quad.\label{eqpot}
\eq
In order to obtain the force of a
spherical halo correctly in the thin disc approximation,
 we use a special approximation
(see subsection \ref{halo}). The relative
contribution of the stellar, the gaseous, and the DM-component to the surface
density (up to $|z|=z_\mathrm{max}$) are given  by the input parameters
$Q_\mathrm{s},Q_\mathrm{g},Q_\mathrm{h}$.

\subsection{Stellar disc \label{stars}}

We assume that the disc of NGC 5907 is, like the thin
disc of the Milky Way, composed of a sequence of stellar subpopulations with
increasing vertical scale height with age. This can be parametrised by the star
formation history and the dynamical heating function. For convenience we
use here the normalisation to the central profile at $R=0$.
The stellar component is composed of a sequence of isothermal subpopulations
characterised by the IMF, the chemical enrichment $[Fe/H](t)$, 
the star formation history $SFR(t_\mathrm{a}-t)$, and the dynamical evolution described by
the vertical velocity dispersion $\sigma(t)$. Here $t$ is the age of the
subpopulation running back in time from the present time $t_\mathrm{a}=12$\,Gyr
(which is the adopted age of the disc). 
We include mass loss due to stellar evolution and retain the
stellar-dynamical mass fraction $g(t)$ (stars + remnants) only. The mass lost
by stellar winds, supernovae and planetary nebulae is mixed implicitly to the
gas component. 

With the Jeans equation the vertical
distribution of each isothermal subpopulation is given by
\bq
\rho_\mathrm{s,j}(z)=\rho_\mathrm{s0,j}\exp\left( \frac{-\Phi(z)}{\sigma^2(t_\mathrm{j})}\right)
\quad,
\eq
where $\rho_\mathrm{s,j}$ is actually a 'density rate', the density per age bin. The
connection to the $SFR$ is given by the integral over $z$
\bq
g(t_\mathrm{j})SFR(t_\mathrm{a}-t_\mathrm{j})=\int_\mathrm{-\infty}^{\infty}\rho_\mathrm{s,j}(z)\dd z\quad.
\eq
The (half-)thickness $h_\mathrm{p}(t_\mathrm{j})$ is defined by
the midplane density $\rho_\mathrm{s0,j}$ through
\bq
\rho_\mathrm{s0,j}=\frac{SFR(t_\mathrm{a}-t_\mathrm{j})}{2h_\mathrm{p}(t_\mathrm{j})}\quad .
\eq
The total stellar density is
\bq
\rho_\mathrm{s}(z)=\int_0^{t_\mathrm{a}}\rho_\mathrm{s,j}(z)\dd t \quad,
\eq
which then determines the potential $\Phi_\mathrm{s}(z)$ via
the Poisson Eq. (\ref{eqpoi}). The stellar surface density $\Sigma_\mathrm{s}$ is
connected to the integrated star formation $S_0$ by the effective
stellar-dynamical fraction $g_\mathrm{eff}$
\bq
\Sigma_\mathrm{s}=\int\rho_\mathrm{s}(z)\dd z=g_\mathrm{eff}S_0\label{eqSigmas}
\eq
with
\bq
S_0=\int SFR\, \dd t\quad \mbox{and}\quad 
        g_\mathrm{eff} =\frac{\int g(t)SFR(t_\mathrm{a}-t)\dd t}{S_0}\quad .\label{eqS0}
\eq
The effective scale height $z_\mathrm{s}$ of the stellar disc is connected to the maximum
velocity dispersion $\sigma_\mathrm{e}$ of the subpopulations and the total surface
density $\Sigma_\mathrm{tot}$ via
\bq
z_\mathrm{s}=C_\mathrm{z}z_\mathrm{e}=C_\mathrm{z}\frac{\sigma_\mathrm{e}^2}{2\pi G \Sigma_\mathrm{tot}}\quad, \label{eqze}
\eq
where $z_\mathrm{e}$ is the scale height of an isothermal component above a disc
with total surface density $\Sigma_\mathrm{tot}$. The 
shape correction factor $C_\mathrm{z}$ is of order unity and is determined at
 $z=(3\pm 0.5)\,z_\mathrm{e}$.

\begin{figure}[t]
\centerline{
  \resizebox{0.98\hsize}{!}{\includegraphics[angle=270]{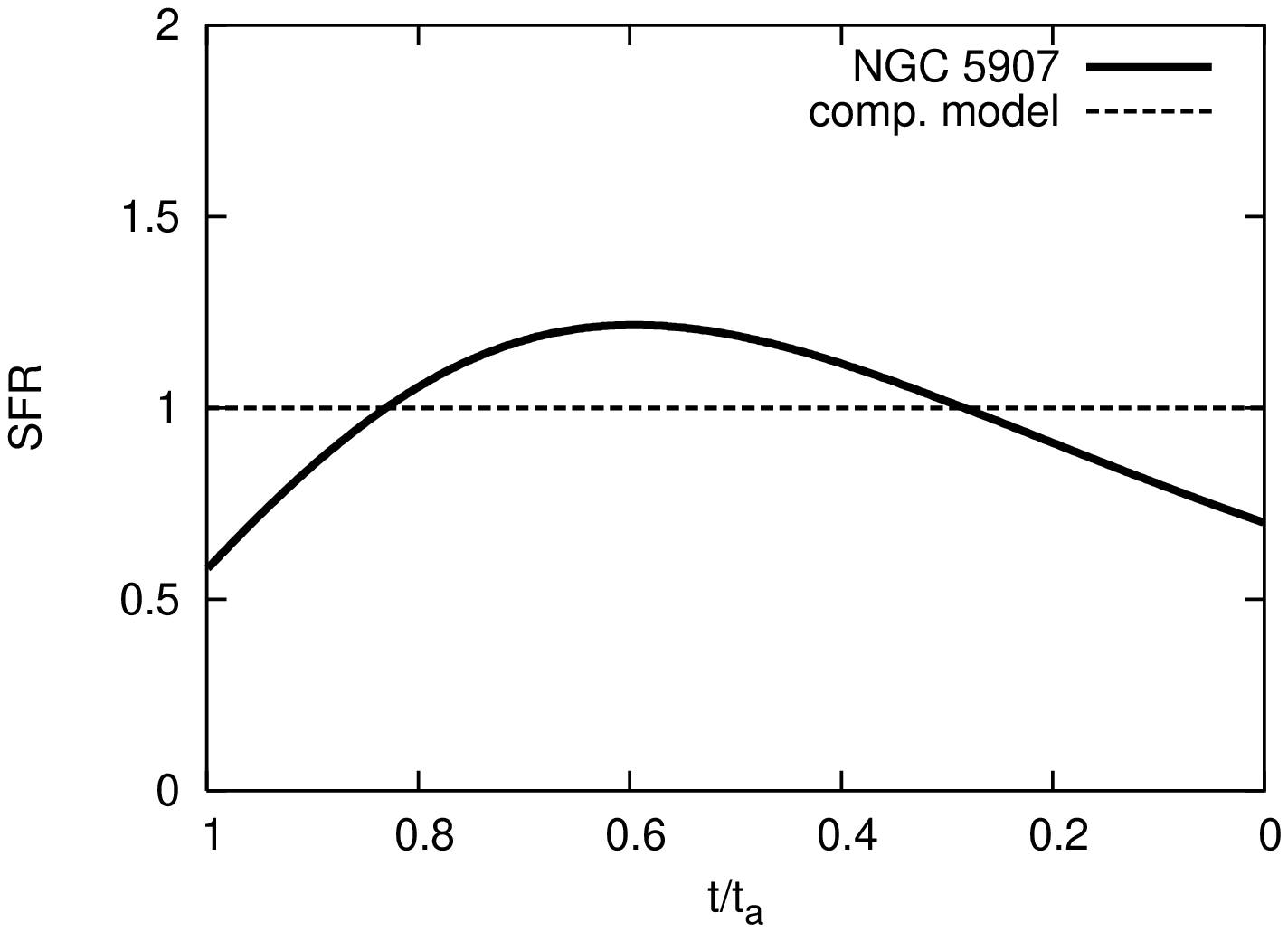}}
  }
\centerline{
  \resizebox{0.98\hsize}{!}{\includegraphics[angle=270]{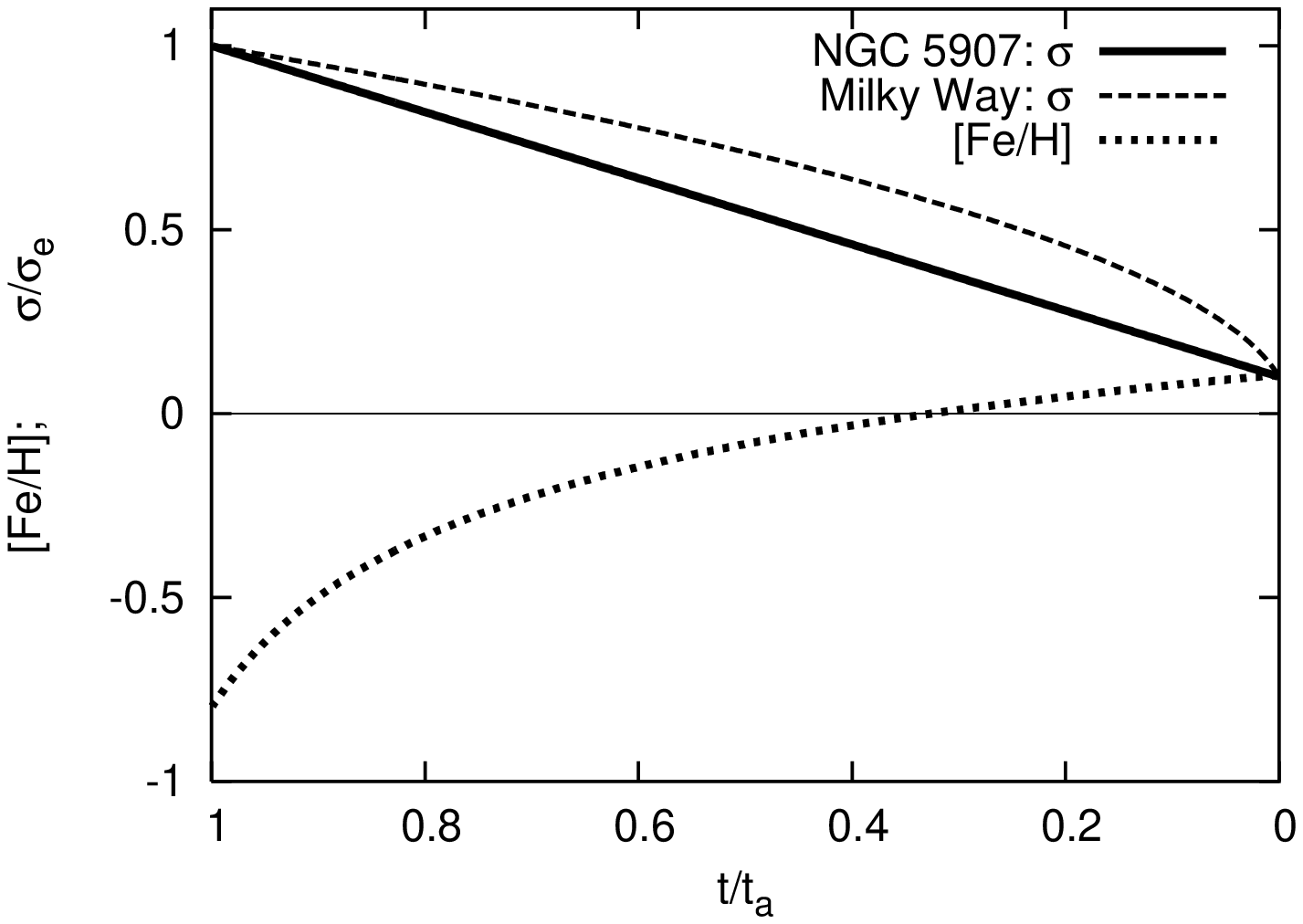}}
  }
\caption[]{
The upper panel shows the normalised $SFR/S_0$ as function of normalised age for the
final model of NGC 5907. The dashed line is the constant SFR of a comparison
model. The lower panel gives the corresponding heating function 
$\sigma(t)/\sigma_\mathrm{e}$
normalised to the final velocity dispersion $\sigma_\mathrm{e}$ (full line)
and the chemical
enrichment (dotted line). 
The dashed line is the heating function of the solar neighbourhood
used for the comparison model.
}
\label{figsfr}
\end{figure}

The metalicity $Z$ affects the stellar lifetimes, luminosities and colours of the
subpopulations. $Z$ shows a vertical gradient, since the mean age of the stellar
population is correlated with the vertical distribution.
In order to account for the systematic influence of the metal enrichment
we adopt a moderate metal enrichment similar to the solar neighbourhood
(Twarog \cite{twa80}, Edvardsson et al. \cite{edv93}). We
use the enrichment law for oxygen 
\bq
Z=z_\mathrm{s}+Y\ln\left(1+a\frac{t}{t_\mathrm{a}}\right)
        \quad\mbox{with}\quad Y=\frac{Z_\mathrm{p}-z_\mathrm{s}}{\ln(1+a)}
\eq
with $Z$ normalised to the solar abundance and the conversion law
$[Fe/H]=2[O/H]$
as  a simple analytic description to account for the enrichment delay by
SN1a. This model
is borrowed from a closed-box model with an $n=2$ Schmidt-law for the star
formation (Lynden-Bell \cite{lyn75}, Just et al. \cite{jus96}).
We use $a=5$, initial and present metalicity $z_\mathrm{s}=0.4$, $Z_\mathrm{p}=1.13$
corresponding to $[Fe/H]=-0.8$ and $[Fe/H]=0.1$ with solar metalicity
$Z_\mathrm{\sun}=0.018$, respectively (see Fig. \ref{figsfr}). 

In the fitting procedure (Sect. \ref{discparam}) a pair of star formation
history and heating function is selected to derive the intrinsic structure of
the disc.
The star formation history and heating function of the final model are
shown in Fig. \ref{figsfr}.

\subsection{Gas and dust component \label{gas}}

For the gas and dust component we use simple models to account for the
gravitation of the gas and the extinction of the dust. In the radial direction
we adopt an exponential profile with  scale length 
$R_\mathrm{d}=q_\mathrm{d}R_\mathrm{s}$, where
we allow for a difference in the scale length of the stellar component by the
factor $q_\mathrm{d}$.
For the extinction
by the dust component we use a simple exponential profile with vertical
scale height $z_\mathrm{d}$. 

The vertical profile of the gas component, which is used for the gravitational 
force of the gas, is constructed dynamically 
like the stellar component. The gas distribution is modeled by
distributing the gas with a constant rate over the  velocity
dispersion range $\sigma(t)$ of the young stars up to a maximum age
$t_\mathrm{g}$. By varying $t_\mathrm{g}$ we force the scale height of the gas $z_\mathrm{g}$
 to the same value as that of the dust component $z_\mathrm{d}$. 
 The surface density of the gas
 is related to the stellar surface density by the ratio 
 $Q_\mathrm{g}/Q_\mathrm{s}=\Sigma_\mathrm{g}/\Sigma_\mathrm{s}$, which is determined at the reference
 radius $R_0$ in the fitting procedure (cf Sect. \ref{project}).

\subsection{Dark matter halo \label{halo}}

The halo does not fulfil the thin disc approximation. For a spherical halo we
get the vertical component of the force to lowest order from
\bq
\frac{\dd\Phi_\mathrm{h}}{\dd z}= \frac{GM_\mathrm{R}}{R^2}\frac{z}{R}\quad, 
\eq
with $r^2=R^2+z^2$ and $M_\mathrm{R}$ is the enclosed halo mass inside radius $R$. 
Comparing this with the one-dimensional Poisson equation from the thin disc
approximation (Eq. \ref{eqpoi} integrated over $z$ 
near the midplane to lowest order for small $z$)
\bq
\frac{\dd\Phi}{\dd z}= 4\pi G \rho_0 z
\eq
we should use for the local halo density 
\bq
\rho_\mathrm{h0}=\frac{M_\mathrm{R}}{4\pi R^3}
\eq
which exactly corresponds to the singular
isothermal sphere. Therefore we can
 use the thin disc approximation also for the halo, if we use the local halo 
 density $\rho_\mathrm{h0}$ and the halo velocity dispersion 
 $\sigma_\mathrm{h}$ estimated from the rotation curve
 by adopting an isothermal spherical halo. The effect of
 a cored halo, anisotropy and flattening is neglected here. 
 For other halo profiles correction factors would
be necessary introducing some inconsistency in the halo profile description and
leading to a different local halo density. The latter would be more important, 
because the effect of the halo potential on the disc is stronger than the
adiabatic contraction of the halo in the disc potential. 

\begin{figure*}[t]
\centerline{
  \resizebox{0.98\hsize}{!}{\includegraphics[angle=0]{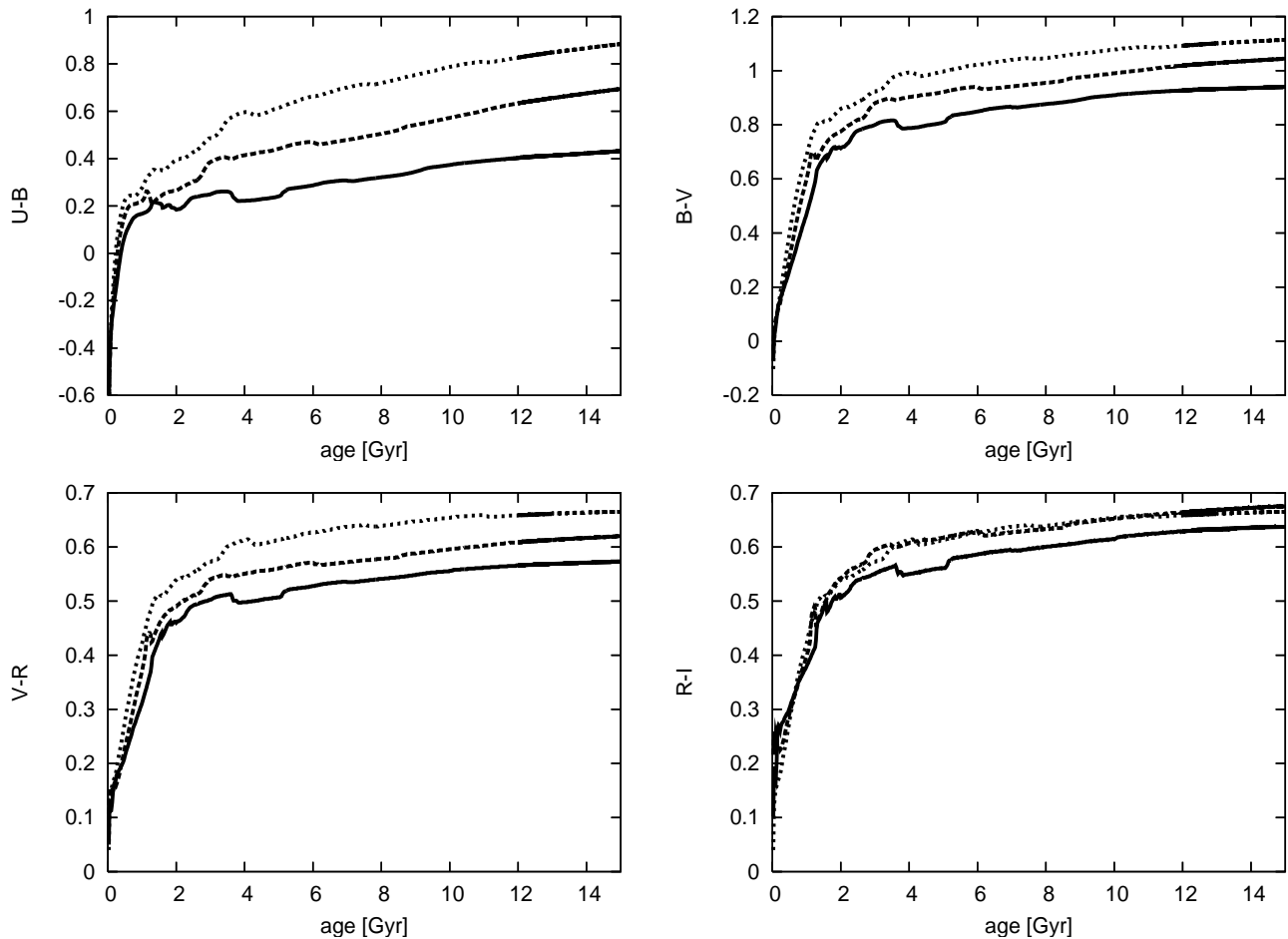}}
  }
\caption[]{
colour index evolution of single stellar populations modelled with the PEGASE code
(Fioc \& Rocca-Volmerange \cite{pegase}). These SSPs are created in a star-burst
lasting for 25 Myr. We use a Scalo IMF (Scalo \cite{sca86}). Three different 
metalicities are shown in the plots: solid: Z = 0.008, long-dashed: 
Z = 0.02, short-dashed: Z = 0.04. 
}
\label{colour_evolution}
\end{figure*}

\subsection{Radial structure \label{radial}}

The parameters of the self-consistent vertical profile described in the last
 section are determined at a scaling radius $R_0=10$\,kpc. 
 The determination of the disc properties at $R_0$ are
 described in Sect. \ref{discparam}. For the radial structure of the disc we use a
 simple exponential extension up to a cutoff radius $R_\mathrm{max}$. 
 We do not account for the radial variation of intrinsic disc 
properties, which are 1) the star formation history or metalicity to fit
the radial colour index gradients, 2) the increasing dark matter mass fraction
with radius, 3) the possible variation of the star/gas surface density ratio
due to different radial scale lengths, 4) the
breakdown of the thin disc approximation in the
innermost part of the disc, 5) the bulge potential and luminosity in the inner
region.

The stellar disc model is then given by
\bq
\rho_\mathrm{s}(R,z)=\rho_\mathrm{s}(z)\exp(-R/R_\mathrm{s}) \qquad\mbox{for}\quad R<R_\mathrm{max}
\label{eqrad}
\eq
with radial scale length $R_\mathrm{s}$, cutoff radius $R_\mathrm{max}$ and constant scale
height $z_\mathrm{s}$.

For the gas and dust component we allow for a different radial scale length
$R_\mathrm{g}=R_\mathrm{d}=q_\mathrm{d}R_\mathrm{s}$ with scaling factor $q_\mathrm{d}$. For the conversion of
total masses to density distributions we do not apply a cutoff radius. 
 The corresponding distribution of extinction is
\bq
A_\mathrm{V}(R,z)=A_\mathrm{V,0}
        \exp\left(\frac{-R}{q_\mathrm{d}R_\mathrm{s}}-\frac{|z|}{z_\mathrm{d}}\right) 
\label{eqdust}
\eq
with central extinction coefficient $A_\mathrm{V,0}$.

\subsection{Stellar population synthesis \label{pop}}

The intrinsic luminosities in the different bands are determined by stellar
luminosity synthesis.
We use the stellar population synthesis code PEGASE 
(Fioc \& Rocca-Volmerange \cite{pegase}) to produce ``pseudo'' simple stellar 
populations (SSPs), which are used for computing the intrinsic
luminosity distribution of the stellar disc. This means that the PEGASE code 
is used to calculate the
integrated colour indices for a stellar population created in a single star-burst at 
different time-steps. These SSPs are then used to assemble
a stellar population with a given star formation history, in the sense that
the star formation history is assembled by a series of star-bursts. 
Since we do the assembling of SSPs ourselves, we have 
to take into account our own treatment of chemical enrichment. 

Our ``pseudo'' SSPs are modeled by a constant star formation rate with a
duration of 25 Myr. As a input parameter for the PEGASE code we use a fixed 
Scalo-like IMF (Scalo \cite{sca86}) given by
\bqn
\dd N &\propto& M^{-\alpha}\dd M\\
&& \alpha=\left\{\begin{array}{lcc}
        1.25&&0.08\le M/\msun < 1\\
        2.35&for&1\le M/\msun < 2\\
        3.0&&2\le M/\msun < 100
        \end{array}\right.
\eqn

The stellar mass of this population is normalised to 
$10^{10}$ solar masses. We use the PEGASE code to produce flux tables
of broadband colours at linear spaced time-steps of 25 Myr. This is done
for the three different metalicities ($Z=0.008; 0.02; 0.04$), which are
the input parameters into our code, where the chemical enrichment is
approximated by linear interpolation of populations with these fixed 
metalicities. 

Fig. \ref{colour_evolution} shows the colour index evolution of our SSPs for the 
three different metalicities. 

\begin{figure}[t]
\centerline{
  \resizebox{0.98\hsize}{!}{\includegraphics[angle=0]{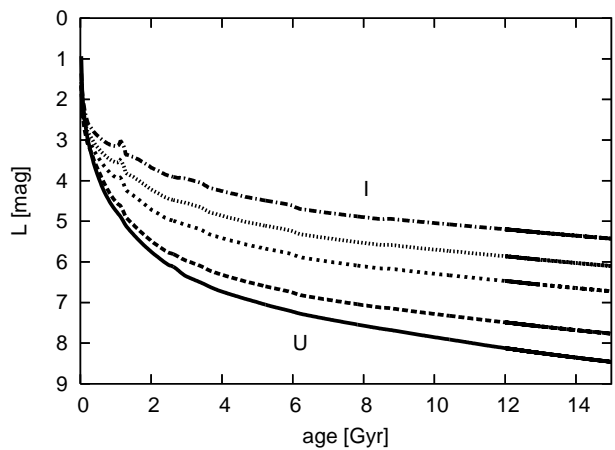}}
}
\caption[]{
Luminosities of single stellar populations modelled with the 
PEGASE code with a Scalo IMF (Scalo \cite{sca86}) for solar metalicities. 
The luminosities are normalised to 1 solar mass. The different lines show
different broadband filters: solid: U, long-dashed: B, short-dashed: V,
dotted: R, dot-dashed: I.
  }
\label{luminosities}
\end{figure}

\begin{figure}[t]
\centerline{
  \resizebox{0.98\hsize}{!}{\includegraphics[angle=0]{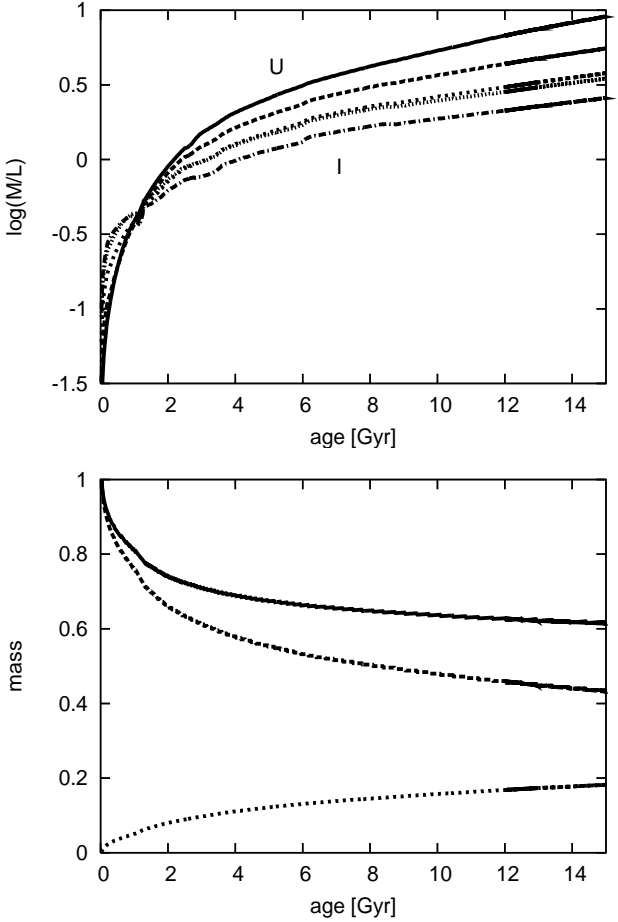}}
}
\caption[]{Evolution of the
stellar mass-to-light ratios (including remnants) and of the mass of 
single stellar populations 
modelled with the PEGASE code with a Scalo IMF (Scalo \cite{sca86}). 
The upper plot shows the 
M/L evolutions of a solar metalicity SSP in different broadband filters: 
solid: U, long-dashed: B, short-dashed: V, dotted: R, dot-dashed: I. 
We use the solar luminosities U = 5.54, B = 5.38, V = 4.75, R = 4.22, 
I = 3.87. The lower plot shows the mass evolution of the stellar population,
normalised to its initial mass. The solid line shows the
total stellar mass, and the other lines its two components separately:
long-dashed: luminous stars, short-dashed: stellar remnants.
  }
\label{masses}
\end{figure}

The normalised luminosities in different
broadband filters are shown in Fig. \ref{luminosities}, and the mass-to-light
ratios are shown in Figure \ref{masses}. These SSP lookup-tables 
are used to assemble the total fluxes and stellar masses for a given star 
formation history. In Fig. \ref{masses} the fraction of initial mass in luminous
stars, remnants and the sum of both, which goes into the stellar disc mass and
the stellar mass-to-light ratio, is plotted for solar metalicity.

\section{Radiative transfer and the fitting procedure
 \label{project}}

Here we describe the computation of the surface density profiles and explain the
main effects on the vertical profiles due to the variation of 
the disc parameters.

The disc model is derived by performing a three-step iteration cycle until good
agreement of the model profiles with the observed surface-brightness and colour index 
profiles is reached. 
In the first step the normalised vertical disc profile is derived by adopting a
star formation history and dynamical heating function combined with the relative
contribution of gas and DM halo component. The intrinsic vertical profile is
constructed to fit the galaxy properties at the scaling radius $R_0$.
In the second step the model is scaled with central surface density, 
radial and vertical scale length of the stellar disc. Then the inclination with
respect to the sky and the dust parameter are chosen to calculate the surface
brightness distribution of vertical cuts
parallel to the minor axis with a radiative transfer code. The geometrical
parameters are varied until a good match to the set of V-band profiles is
reached. Here the scaling of the gas and halo properties are also corrected
iteratively.
In the third step a multi-colour analysis is performed to analyse the
deviations of the model to the complete set of observed surface-brightness and colour index
profiles. These  systematic deviations are used to estimate corrections for the
star formation history and the heating function for an improved model.
In the following these three steps are described in more detail.

\subsection{The normalised vertical profiles \label{discparam}}

We start with a pair $SFR(t_\mathrm{a}-t)$ and $\sigma(t)$ expected for a Sc-galaxy
from a basis set of normalised template functions of different types. For the
calculation of the self-consistent profile we need additionally the
parameters of the gas and halo component normalised to the stellar parameter.
These are 
\\1) the fractional surface densities of gas $Q_\mathrm{g}$ and DM halo $Q_\mathrm{h}$
(leading to $Q_\mathrm{s}=1-Q_\mathrm{g}-Q_\mathrm{h}$ for the stars) 
\\2) the gas scale height normalised to the scale height of the stars $z_d/z_\mathrm{s}$
determined implicitly by the maximum 'age' of the gas $t_\mathrm{g}$
\\3) the velocity dispersion of the DM halo normalised to the maximum velocity
dispersion of the stars $s_\mathrm{h}=\sigma_\mathrm{h}/\sigma_\mathrm{e}$.

Since  $Q_\mathrm{g}$, $Q_\mathrm{h}$, $s_\mathrm{h}$, and  $t_\mathrm{g}$ depend on the
final scaling of the stellar disc to the observations, they must be corrected
iteratively. To fix the vertical profile we use the disc properties at the
scaling radius $R_0=10$\,kpc. 
For the gas component the surface density is derived from the total gas mass
using the adopted radial scale length. The maximum 'age' $t_\mathrm{g}$ is
chosen to match the vertical scale length of the dust 
$z_\mathrm{g}=z_\mathrm{d}$.
For the dark matter halo we use the mass model
of Sackett et al. (\cite{sac94}) with disc $M/L=1$ in a spherical DM halo.
From the circular velocity of the halo
 $v_\mathrm{c,h}=160$\,km/s and with $2\sigma^2_\mathrm{h}=v^2_\mathrm{c,h}$ the halo velocity 
 dispersion is $\sigma_\mathrm{h}=110$\,km/s and the halo density is 
 $\rho_\mathrm{h0}=v_\mathrm{c,h}^2/(4\pi GR_0^2)=4.7\times 10^{-3}\msun \mathrm{pc}^{-3}$ leading to 
 the surface density 
 $\Sigma_\mathrm{h}=2z_\mathrm{max}\rho_\mathrm{h0}=47.5\,\msun \mathrm{pc}^{-2}$ 
up to a  maximum height of $z_\mathrm{max}=5.1$\,kpc.
The initial values for the stellar disc are also taken from the total stellar
mass in Sackett et al. (\cite{sac94}) and then iterated. 
The normalisation of the velocity
dispersions are given by $\sigma_\mathrm{e}$ also determined iteratively 
from Eq. \ref{eqze}.  

The calculation of the self-consistent profile includes the
intrinsic luminosity profiles in all bands using stellar
population synthesis.

\subsection{Extinction and projection to the sky \label{sky}}

The next step is to choose the stellar scale height $z_\mathrm{s}$ (Eq. \ref{eqze}), the
radial scale lengths $R_\mathrm{s}$ and the cutoff radius $R_\mathrm{max}$ 
(Eq. \ref{eqrad}), 
the central surface density of stars $S_0$ (Eqs. \ref{eqSigmas}, \ref{eqS0})
and the dust properties (Eq. \ref{eqdust}) including the inclination $i$ of the
disc.

With these parameters the projection of the galaxy to the sky is calculated. We
use a simple radiative transfer code neglecting scattered light. 
Some test runs performed by M. Xilouris has shown
that scattered light contributes only a few percent to the surface brightness
of typical edge-on galaxies. This is different to the significant influence of
scattered light on the determination of the dust temperature in the disc
(Misiriotis et al. \cite{mis01}). With the $x$-axis along the line of sight,
$y$-axis along the major axis, and $z$-axis along the minor axis we integrate
for a set of vertical cuts highly resolved $z$-profiles of surface
brightness profiles in U, B, V, R, and I. The resolution is $dx\sin(i)=125$\,pc
along the line of sight and $dz\sin(i)=12$\,pc parallel to the minor axis and we
restrict the integration to the cylinder bounded by $R_\mathrm{max}$ and $z_\mathrm{max}$.
Possible foreground extinction due to dust outside $R_\mathrm{max}$ is neglected.

The extinction and reddening features of the vertical profiles are very
sensitive to the dust geometry and the inclination (for a detailed discussion
see Just et al. \cite{jus96}). Therefore we test a large range of parameters to
find the best match to the observed profiles.

Even if the match of the V-band profiles is not satisfactory we proceed to the
multi-colour analysis to determine how to change the 
star formation history and/or the heating function.

\subsection{Multi-colour analysis \label{fitting}}

Here we give an overview over the main effects on the vertical profiles due to 
the variation of the disc parameters. 
For details see Just et al. (\cite{jus96}).
The global parameters of the stellar disc are determined
by the surface brightness distribution at large heights $z$, where extinction
can be neglected. The vertical cuts near the midplane are influenced by the dust
distribution and the detailed distribution of the stellar subpopulations of
different age. 

We first discuss the influence of the dust distribution. 
If the dust scale height is larger than the scale height of the intrinsic star
light near the midplane, then extinction leads to a flattening of the profile in
the centre but no minimum (dust lane). Increasing the inclination results in a
smearing out of the dust feature. Increasing the dust mass leads to a deeper and
wider dust lane in the surface-brightness profiles but only a broadening in the colour index
index profiles; also the radial extension becomes larger. 
With a radial scale length larger than the stellar scale length the extinction
feature is distributed more shallowly in the radial direction and the dust lane shows
a larger offset to the major axis.

The systematic variations of the shape of the vertical profiles from red to blue
are not only a signature of reddening near the midplane, but are also due to the
relative distribution of young and old stars. 
Changing the star formation history leads to a different
fraction of luminous young blue subpopulations. A higher fraction of young and
intermediate age populations with O,B and A-stars gives an excess in the
surface-brightness profiles near the midplane which cannot be suppressed fully by dust
extinction (especially in the red colours).
The heating function determines the vertical spread of the subpopulations. Fast
heating for the young populations results in centrally flattened surface
brightness profiles combined with an extended blue regime.

As an example of the differences in the intrinsic luminosity and
colour profiles we show in Fig.~\ref{figlumz} a comparison model with constant
$SFR$ (large fraction of young stars) and the heating function of the solar
neighbourhood (quick heating of young stars).
The effect is a reduced central luminosity  
and a very extended blue regime. 
Therefore less dust is needed in the central region to obtain
the observed surface brightness profiles. 
But the blue wings cannot be hidden by a reliable dust distribution.


\begin{figure*}[t]
\hbox{\hspace{0cm}              
 \psfig{figure=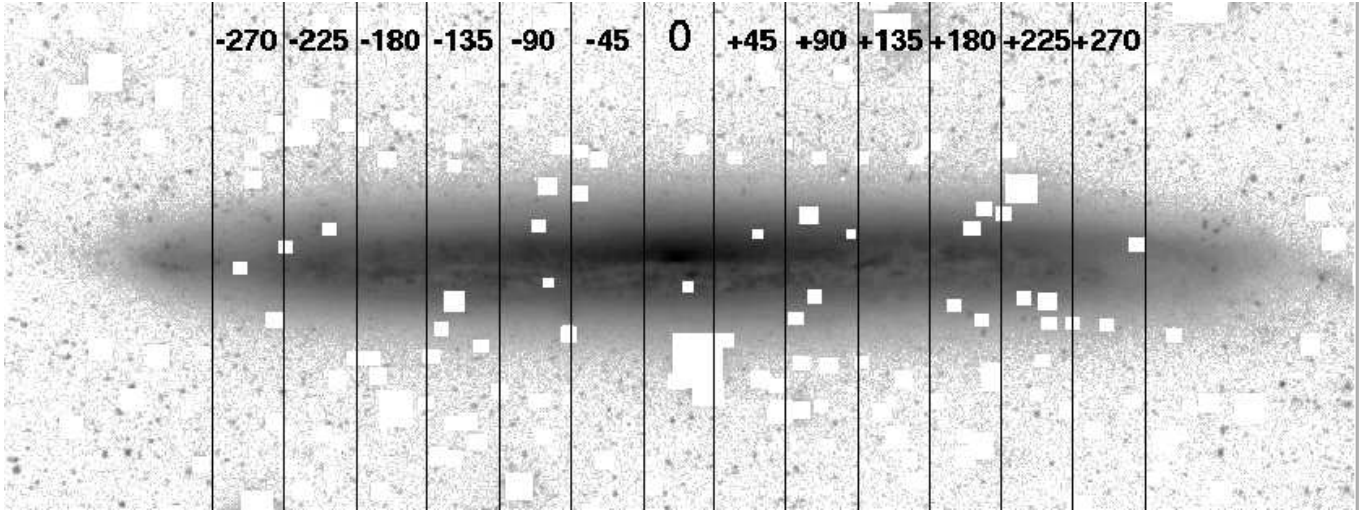,width=18.0cm,clip=}}
\caption[]{Deep V image of NGC 5907 with the stellar mask.
The image is rotated for a horizontal orientation of the galaxy.
The field of view is $850 \times 320$ arcsec. 
The vertical flux profiles were calculated
by mean values over the 13 indicated stripes of 45 arcsec width. The black
lines mark the borders of these stripes. 
The flux values for the each point of the 
profiles were calculated by averaging along horizontal pixel rows of 45 arcsec 
length. The numbers in the stripes give their corresponding central positions
in arcsec. NW is left, SE is right.}
\label{figvprof}
\end{figure*}

\section{Observations and data reduction \label{observ}}

It is important to check the viability of our model by comparison
with multi-colour observational data of NGC 5709. In this section we
describe the observations. The comparison will be presented in 
Sect.~\ref{result}.
  
The observations were performed during July 5 - 11, 1997 using the 2.2m 
telescope of the Calar Alto Observatory, Spain. The CAFOS focal reducer
was used changing the focal ratio of the telescope from f/8 to f/4.4.
The CCD camera was equipped with a SITE 2048 $\times$ 2048 pixel chip. 
The pixel size was 24$\mu$, corresponding to 0.53 arcsec. 
The circular field-of-use had a diameter of 16 arcmin.
Standard Johnson U,B,V and Cousins R,I filters were used.
Several images were exposed in each filter without interrupting the
tracking of the telescope. The exposure times in UBVRI are listed in 
Table~\ref{tablparam}.
 
Sky flats were exposed in every dawn and dusk phase. 
The photometric quality was good or very good in all nights. 
Photometric standard stars ($\approx 40$) in the globular cluster M~92
were exposed every night  
(Christian et al. \cite{christ}, Sandage \& Walker \cite{sand}, 
Cathey \cite{cath},
Stetson \& Harris \cite{stet}).

\begin{table}
\caption[]{The observational data of NGC 5907. Most columns are self 
explaining. $FWHM$ is the full width half maximum of the seeing. 
The two last columns give the total magnitudes and 
errors from our measurements, corrected for galactic foreground extinction.}
\label{tablparam}
\[
\begin{array}{|ccccc|rc|}
\hline      
 Filt    & Date   &    Exp.time     &  Airmass  &  FWHM    & total & phot\\
         & Jul97  &      [min]      &           & [arcsec] & Magn  & error\\
\hline              
 {\rm U} & 11     & 8 \times 20      & 1.10 - 1.70 & 1.5 &  11.567 &\pm 0.070  \\
 {\rm B} &  6     & 8 \times 20      & 1.12 - 1.53 & 1.5 &  11.279 &\pm 0.054 \\
 {\rm V} &  5     & 4 \times 20      & 1.06 - 1.13 & 1.5 &  10.419 &\pm 0.057 \\
 {\rm R} &  5     & 20 + 3 \times 10 & 1.30 - 1.50 & 1.3 &   9.701 &\pm 0.064 \\
 {\rm I} &  5     & 4 \times 10      & 1.17 - 1.26 & 1.8 &   8.763 &\pm 0.084 \\
\hline
\end{array}
\]
\end{table}

\subsection{Data reduction \label{datred}}

The MIDAS program system was used for the data reduction. 
From comparison and evaluation of all bias exposures 
an optimal working bias image was constructed.
Similarly, from comparison and evaluation of all flat-field exposures 
optimal working-flat-field images were constructed for each colour.
Bias subtraction and flat-field division were performed in the usual manner
for each image. The flatness of the sky was checked in each image and corrected 
interactively, if necessary. The flatness of the sky down to less than 0.5\% is 
an important condition for a reliable model of the surface-brightness distribution. 

The images in each colour were added together to a total image.
A rebinning was not necessary since the tracking of the telescope was not 
interrupted between the exposures (this was checked, cf the column $FWHM$ in 
Table~\ref{tablparam}).
By stacking the images in the vertical direction it was possible to 
remove the cosmic pixels during the summation (Kappa-sigma-summation). 
Fig.~\ref{figimagr} shows as an example the total image in R. The faint
filament extending northwards from the left part of the disk is real.
It probably belongs to a remnant of an interacting dwarf galaxy
(Shang et al. \cite{shang}).

The flux calibration was calculated using the standard stars.
The zero-points could be determined with a error between 0.05 (B) to 0.08 (I) mag.
The atmospheric extinction of the corresponding airmass was calculated 
for each single exposure. Each exposure was considered with a corresponding 
weight in the flux calibration of the total image.
The galactic foreground extinction for NGC 5907 is rather small: 
$\rm A_g(B) = 0.01$ taken from RC3 (de Vaucouleurs et al. \cite{dev91}), see also
in Burstein \& Heiles (\cite{burstein}). 
The extinction corrections for the other filters were computed by the
coefficients given in Cardelli et al. (\cite{card}) (their Table 3).
The resulting corrections were 
$\rm A_g = 0.012, 0.010, 0.008, 0.006, 0.004$ mag for U, B, V, R, I, respectively.
Because of the small distance of NGC 5907
the K-corrections are very small and were not considered. 

For the total magnitude of NGC 5907 we obtain the values given in 
Table~\ref{tablparam}.
These values are corrected for galactic foreground extinction.
The errors are between 0.05 and 0.08 mag. The corresponding B value
from the RC3 is $B_\mathrm{tot} = 11.11 \pm 0.10$. Our value is 0.17 mag
weaker. The difference might be explained by the fact that the RC3 value
was obtained by aperture photometry while we have masked the foreground stars in
front of and around the galaxy.

\subsection{Surface-brightness and colour index profiles \label{profile}}

For the comparison with the models we need vertical profiles
of the calibrated flux in UBVRI and the corresponding colour indices.
All arithmetic calculations for this were done with the flux images. The flux
calibrations were done afterwards.

The total images were rotated to set the galaxy into a horizontal position.
A mask with appropriate squares around all fairly bright foreground stars
and background galaxies was constructed from the rotated total V image. 

For the construction of vertical flux profiles
a number of corresponding vertical stripes were defined.
The centre of the first stripe was placed at the centre of the galaxy and 
the others were placed at  
the horizontal positions $\pm 45, \pm 90, \pm135,\pm 180, \pm 225, \pm270$ 
arcsec along the disk. (Fig.~\ref{figvprof}).
This corresponds to 0.5, 1.0, 1.5, 2.0, 2.5, and 3.0 scale lengths.
Each stripe had a width of 45 arcsec, thus the whole disk was covered.
In each stripe we calculated the mean flux along each horizontal 
pixel row of 45 arcsec length. Whenever a stellar mask was hit, this was 
adequately
considered in the mean value. In this way we obtained 13 flux profiles 
over z for each colour U, B, V, R, I. Further on we calculated the average of 
the left and right profiles of the corresponding horizontal positions.

Using our flux-calibration formula these profile 
were transformed into magnitudes. The profiles of the colour indices 
U--B, B--V, V--R, R--I were obtained by subtraction of the
corresponding profiles.

\begin{figure*}[t]
\centerline{
  \resizebox{0.43\hsize}{!}{\includegraphics[angle=270]{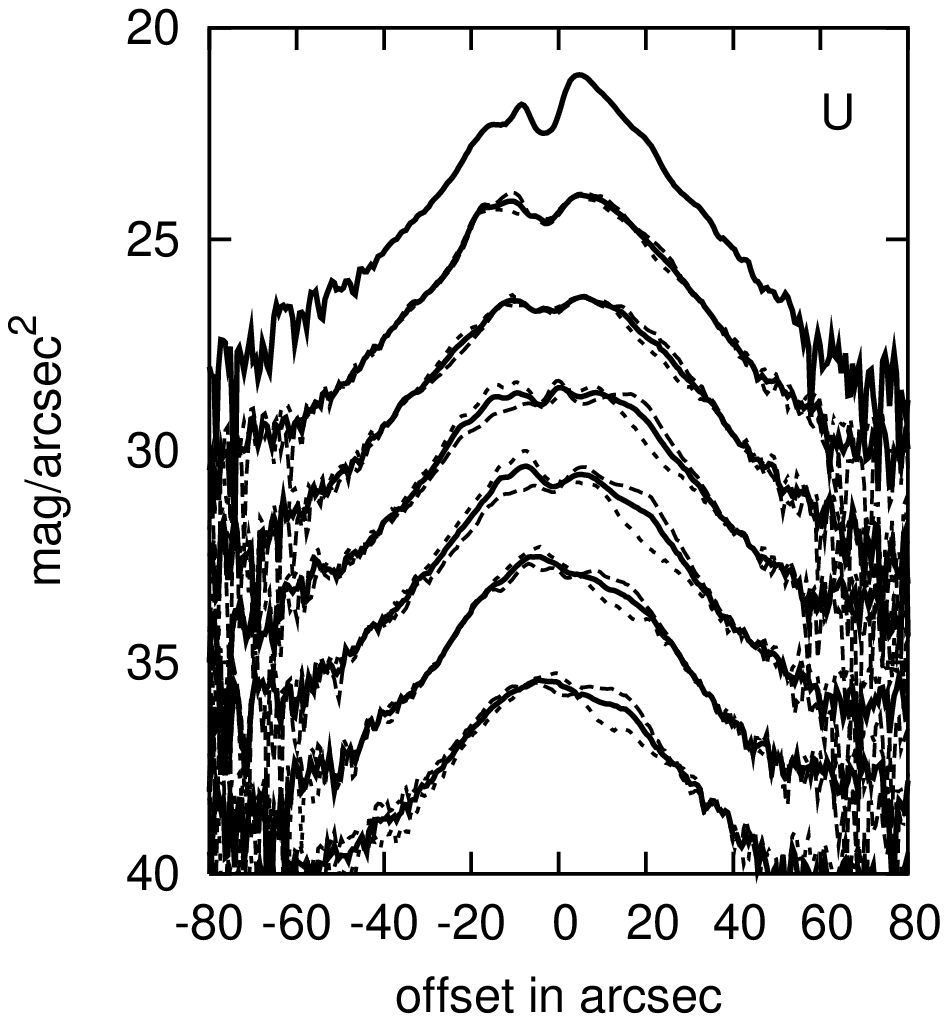}}
  \resizebox{0.43\hsize}{!}{\includegraphics[angle=270]{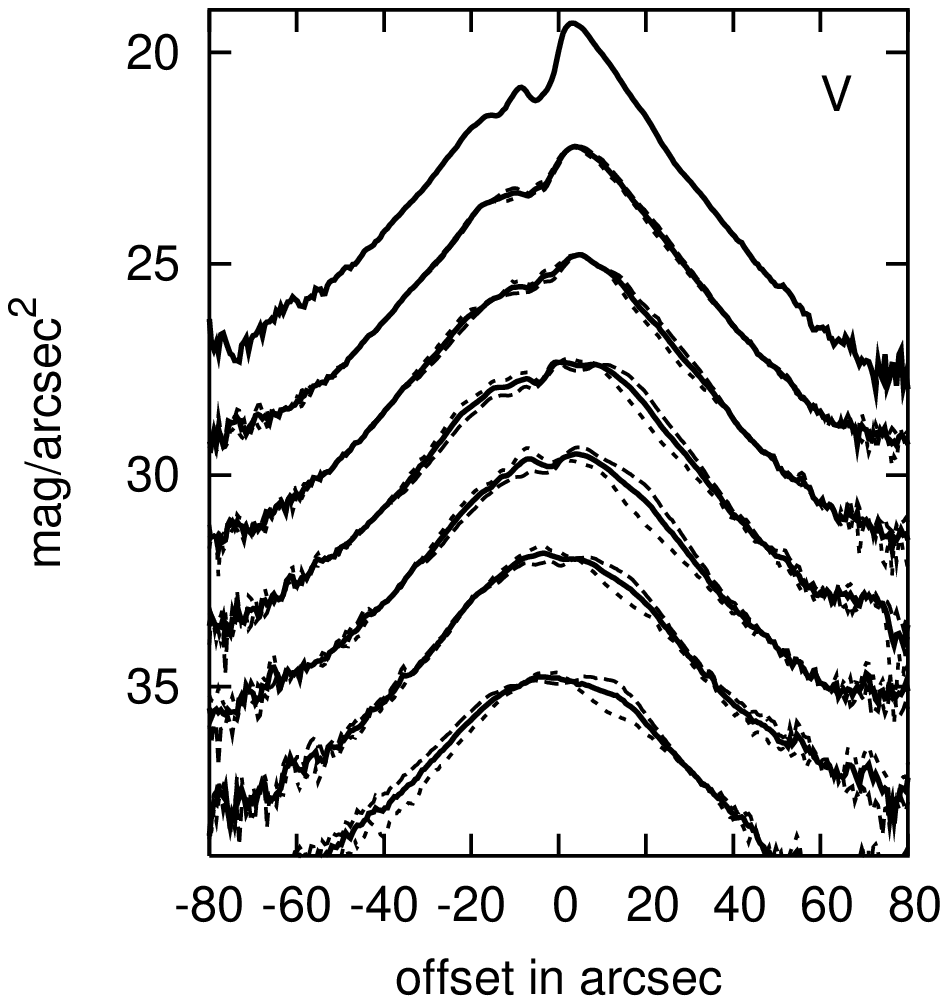}}
  }
\caption[]{
Vertical surface-brightness profiles of NGC 5907 each averaged over 45$\arcsec$ 
in the radial direction with an offset to the minor axis of 
0, 45, 90, 135, 180, 225, and 270$\arcsec$ from top to bottom.
For a better visibility the magnitude scale is shifted by 2 magnitudes
from one to the next.
Long and short dashed lines give the individual profiles for the 
NW(-) and SE(+) sides of the disc and the thick full
lines are the averaged profiles.
}
\label{fig-lumobs}
\end{figure*}

In Fig. \ref{fig-lumobs} we show the vertical cuts in U- and V-band as
examples.
The thin lines are the NW- and SE-profile and the thick lines are the averaged
profiles which we used for the model. The extinction feature (perturbed by the
clumpy structure of the young subpopulation) is clearly visible. In the outer
parts the profiles are exponential and the noise level is well defined. The
small offsets of the profiles in the different profiles show that the warp
discussed in Morrison et al. (\cite{mor94}) seems to be not very systematic or 
is only significant for the young clumpy disc component. 

\begin{figure*}[t]
\centerline{
  \resizebox{0.43\hsize}{!}{\includegraphics[angle=270]{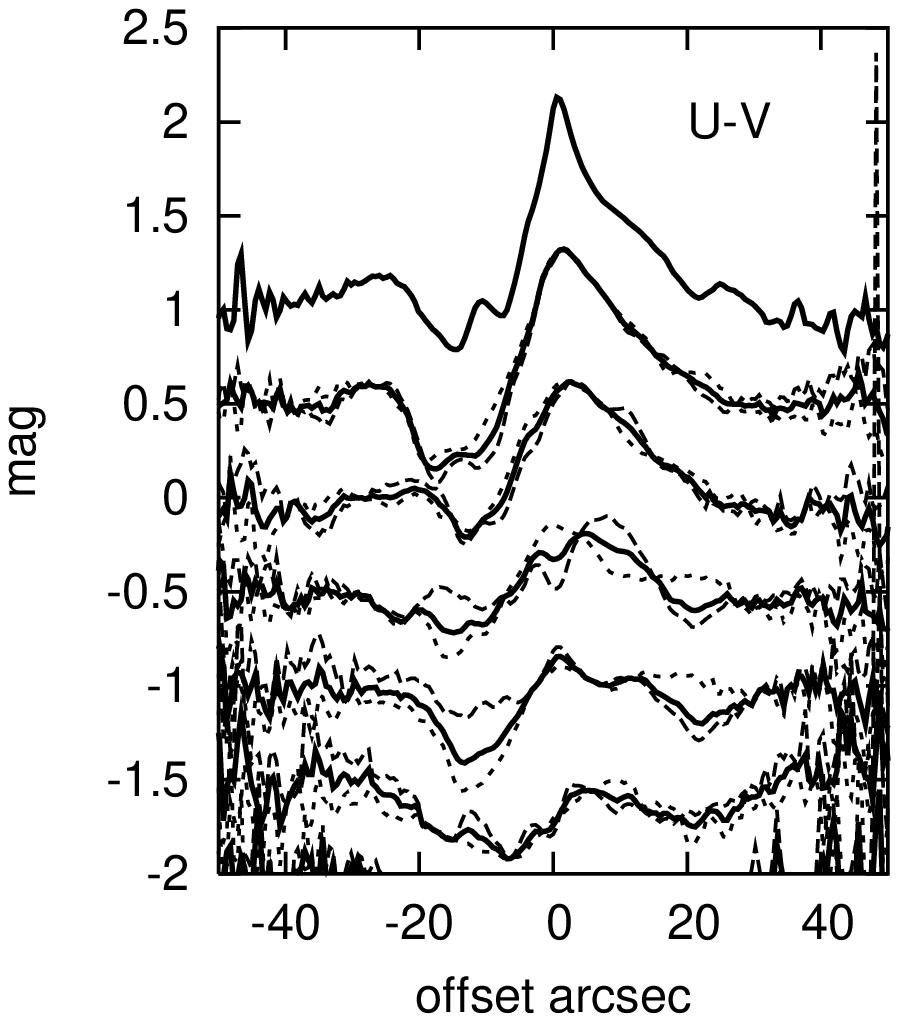}}
  \resizebox{0.43\hsize}{!}{\includegraphics[angle=270]{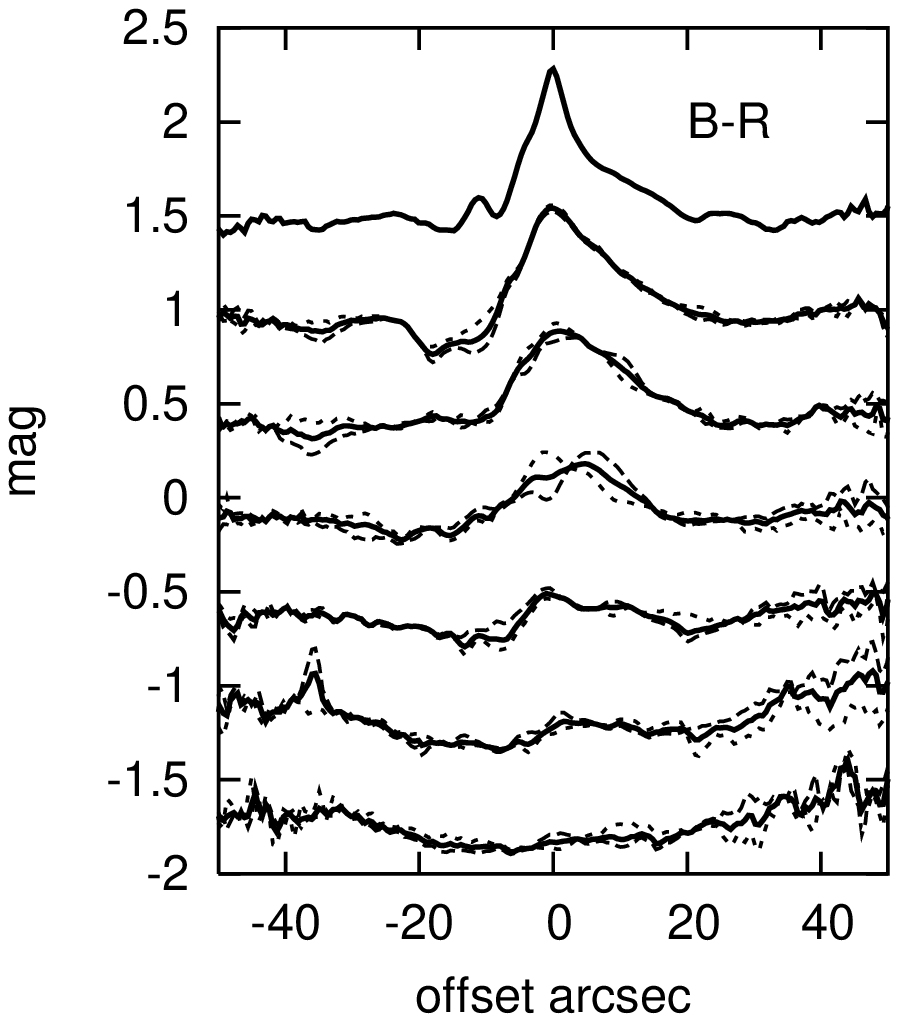}}
  }
\caption[]{
Vertical profiles  of the colour indices U--V and B--R
for NGC 5907 for the same cuts as in Fig.
\ref{fig-lumobs}. From top to bottom, the magnitude scale is shifted in
colour by -0.5 magnitudes from one to the next.
}
\label{fig-colobs}
\end{figure*}

\section{Results \label{result}}

We show the comparison of the final model to the data in
the surface-brightness and colour index profiles. 
Then we give the global parameters of the
disc model, describe the scaling procedure,
and discuss the intrinsic properties of the model.

\begin{figure*}[t]
\centerline{
  \resizebox{0.33\hsize}{!}{\includegraphics[angle=270]{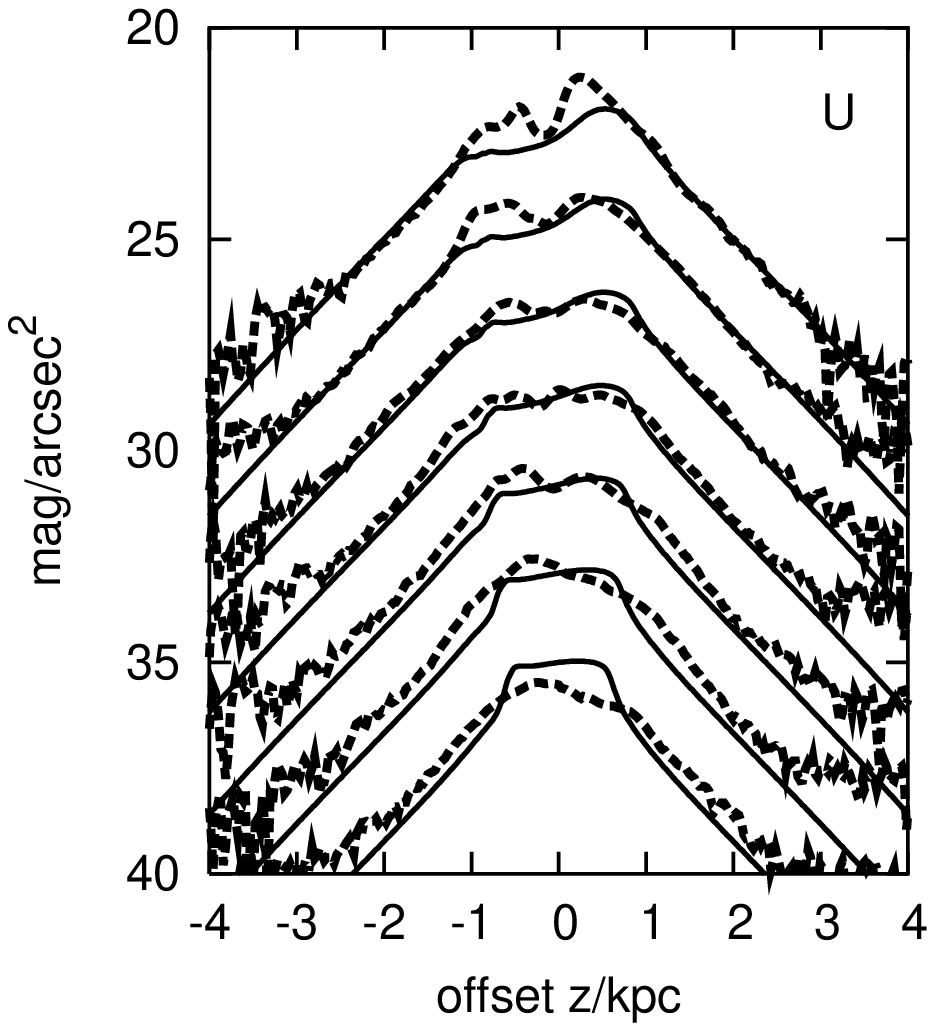}}
  \resizebox{0.33\hsize}{!}{\includegraphics[angle=270]{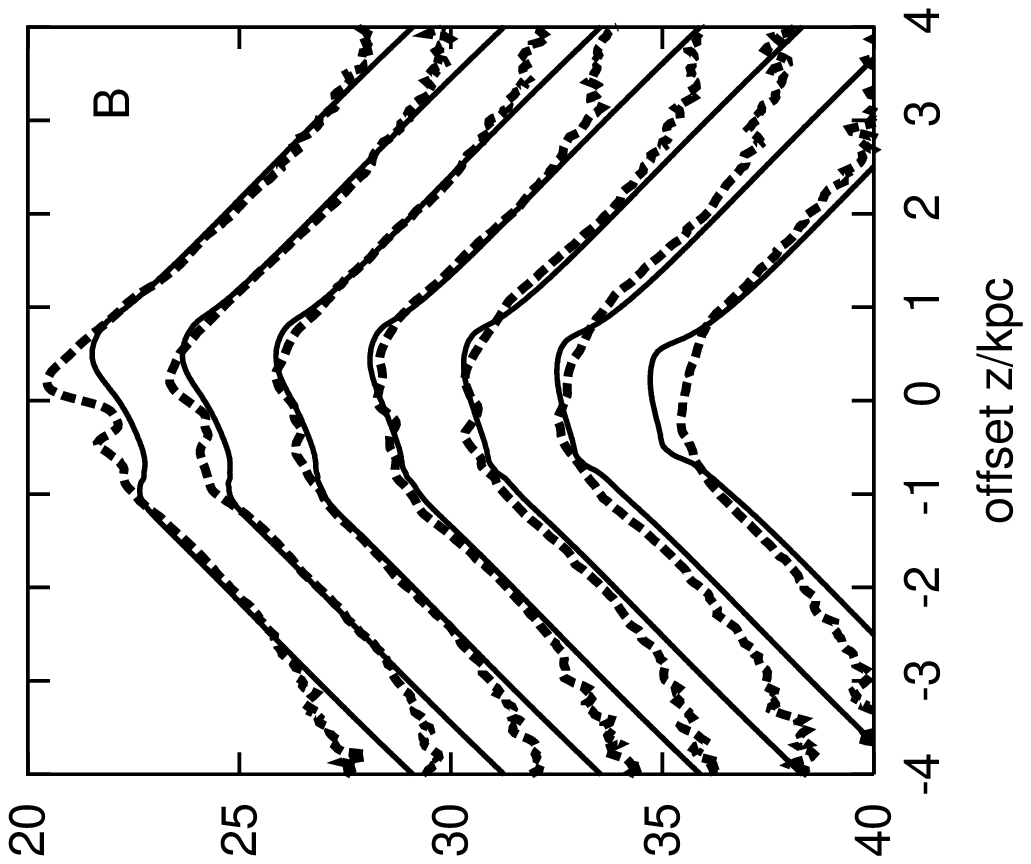}}
  \resizebox{0.33\hsize}{!}{\includegraphics[angle=270]{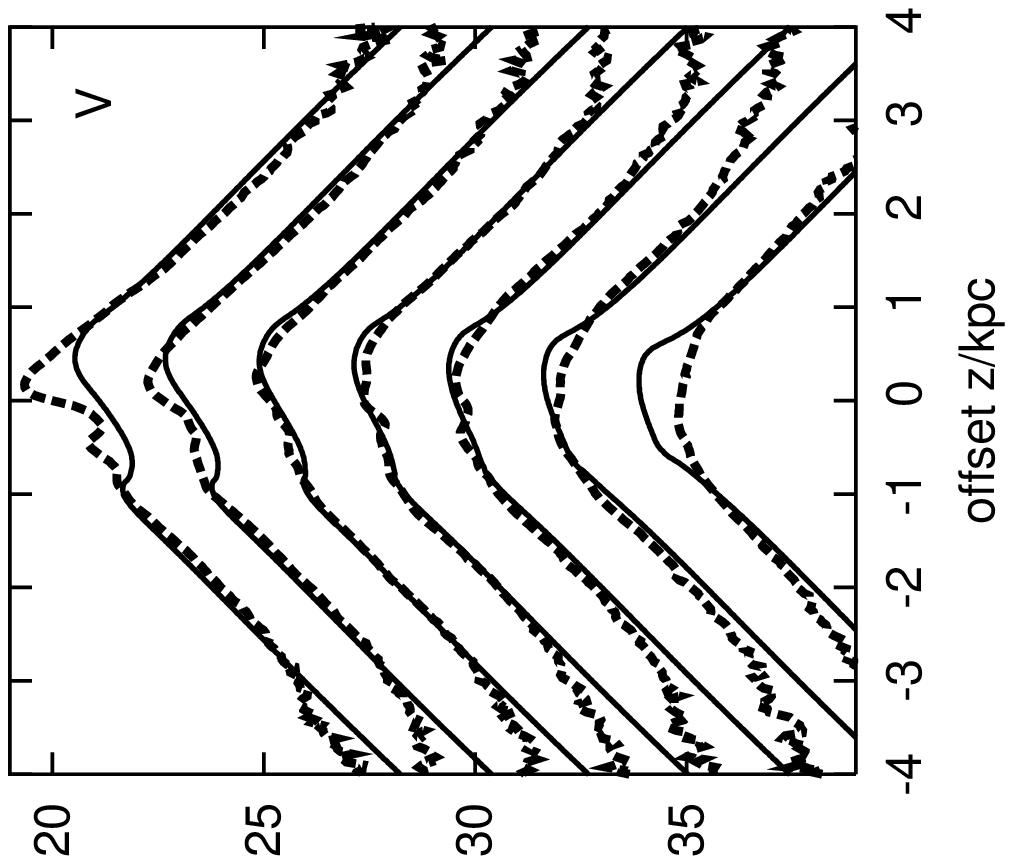}}
  }
\centerline{
  \resizebox{0.33\hsize}{!}{\includegraphics[angle=270]{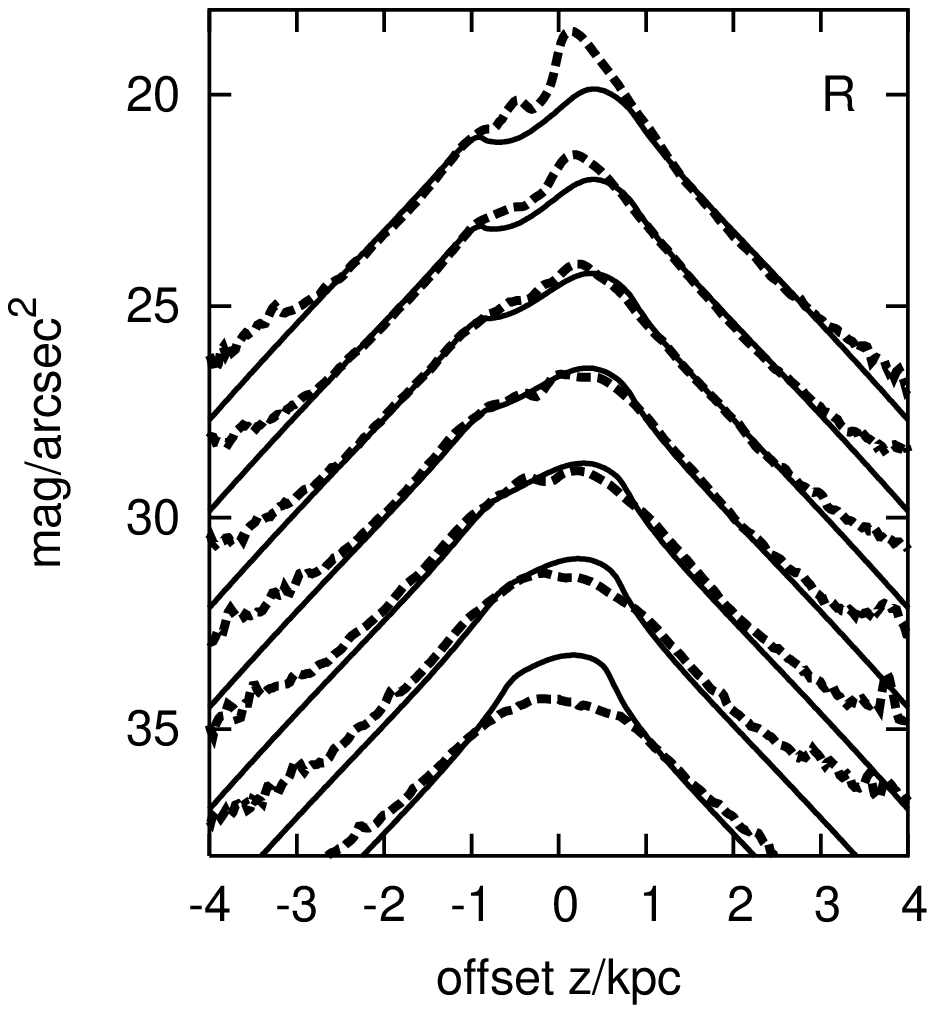}}
  \resizebox{0.33\hsize}{!}{\includegraphics[angle=270]{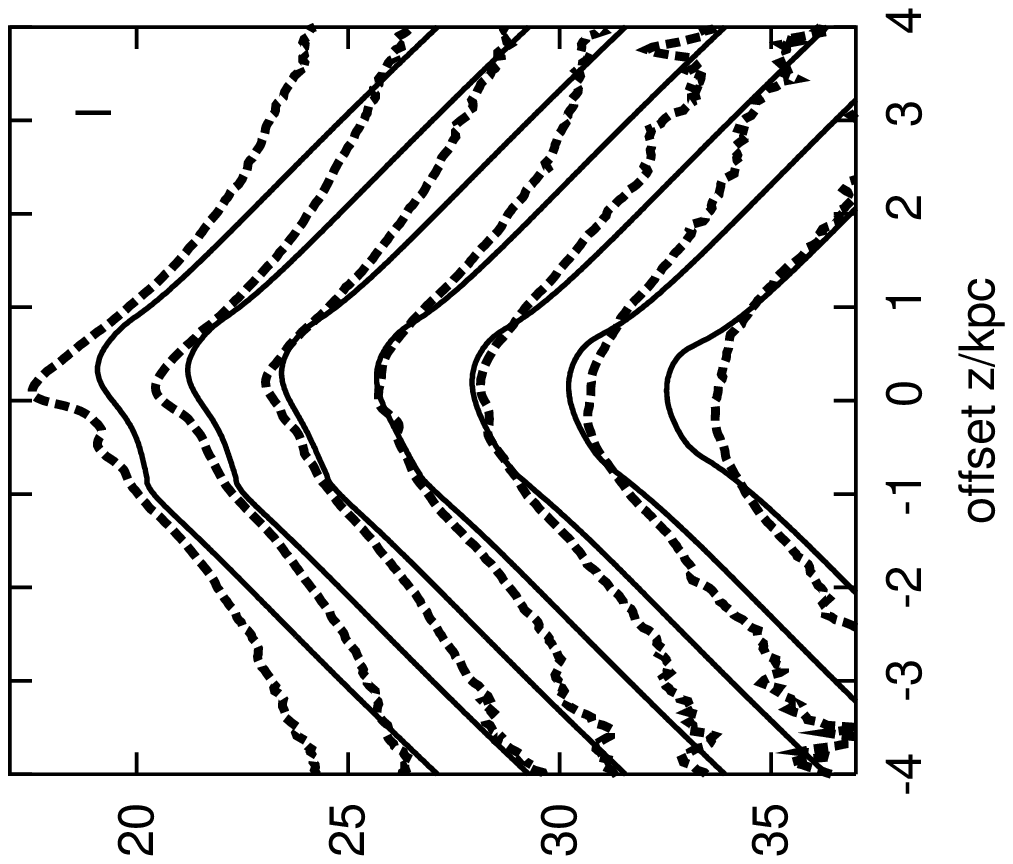}}
  }
\caption[]{Comparison between observations and model.
Vertical surface brightness profiles for NGC 5907 in the U, B, V, R, and I bands
from upper left to lower right. The seven profiles in each plot are
vertical cuts at distances along the major axes 
of 0, 45, 90, 135, 180, 225, and 270$\arcsec$ corresponding to 
0, 2.4, 4.8, 7.2, 9.6, 12.0, and 14.4\,kpc from top to
bottom. Dashed lines represent the data, while solid lines give
the model. From top to bottom, the magnitude scale is shifted downwards in
brightness by 0, 2, 4, 6, 8, 10, and 12 magnitudes, respectively.
}
\label{figlum}
\end{figure*}
\begin{figure*}[t]
\centerline{
  \resizebox{0.33\hsize}{!}{\includegraphics[angle=270]{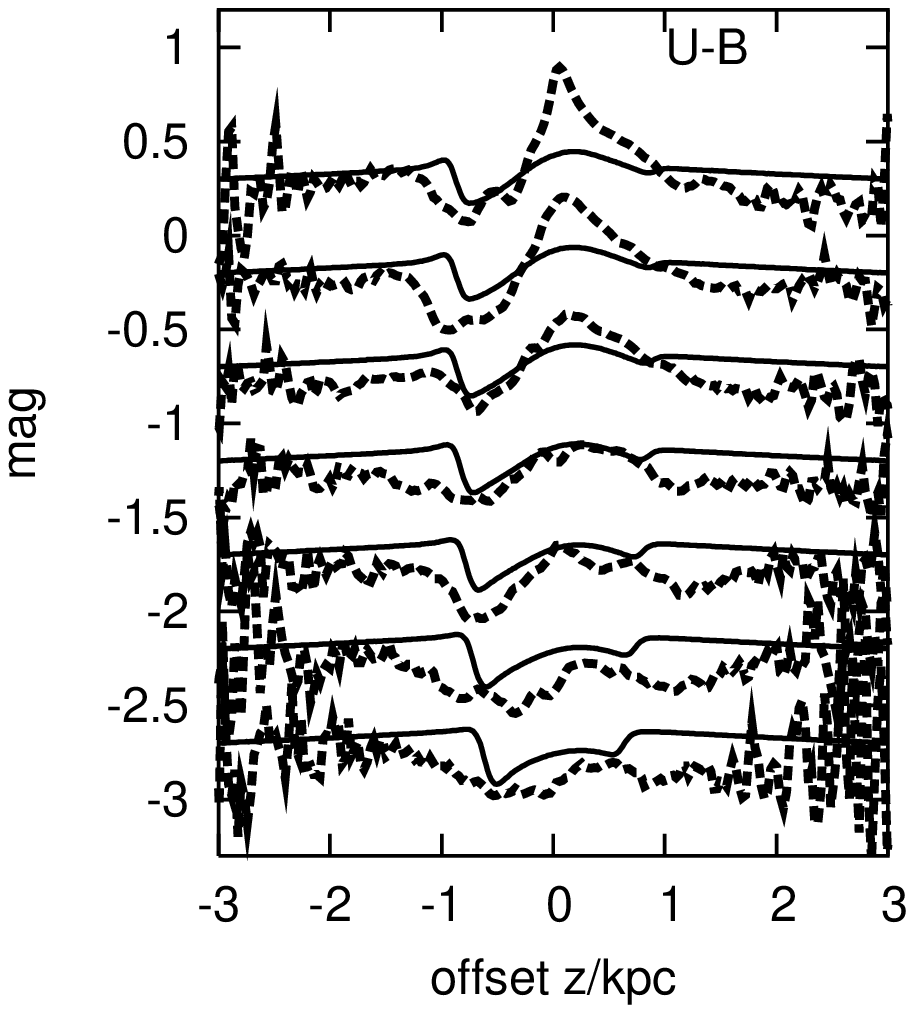}}
  \resizebox{0.33\hsize}{!}{\includegraphics[angle=270]{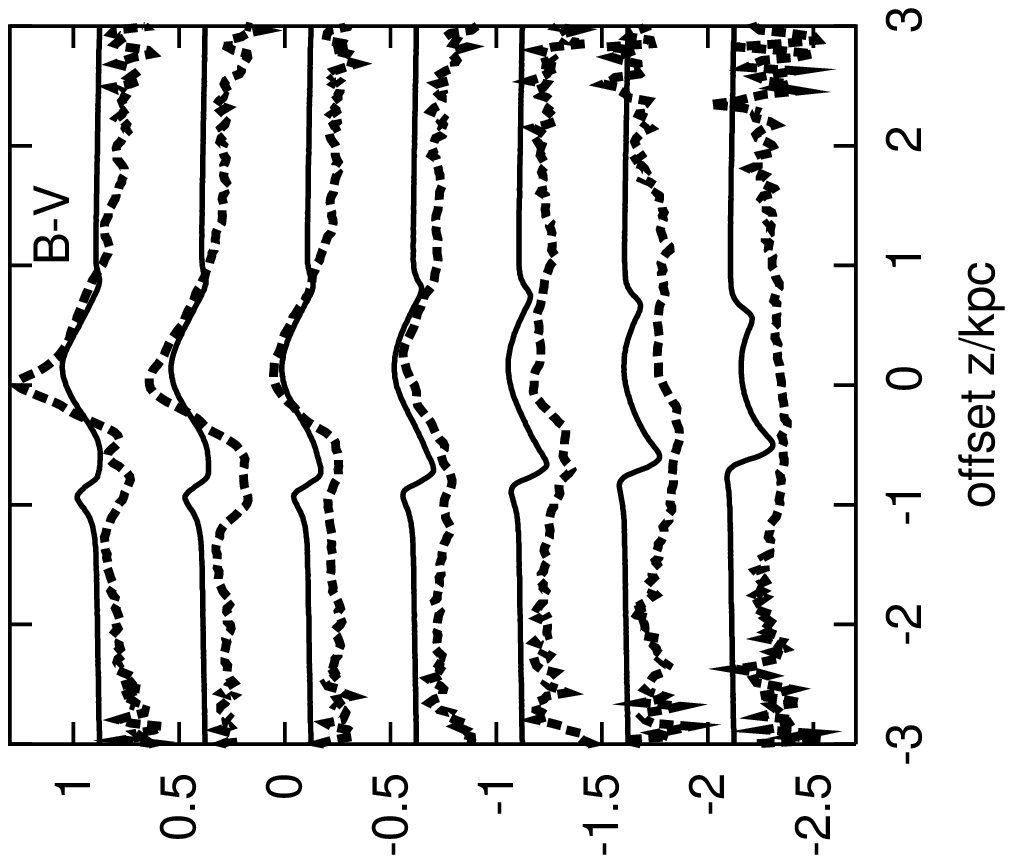}}
  \resizebox{0.33\hsize}{!}{\includegraphics[angle=270]{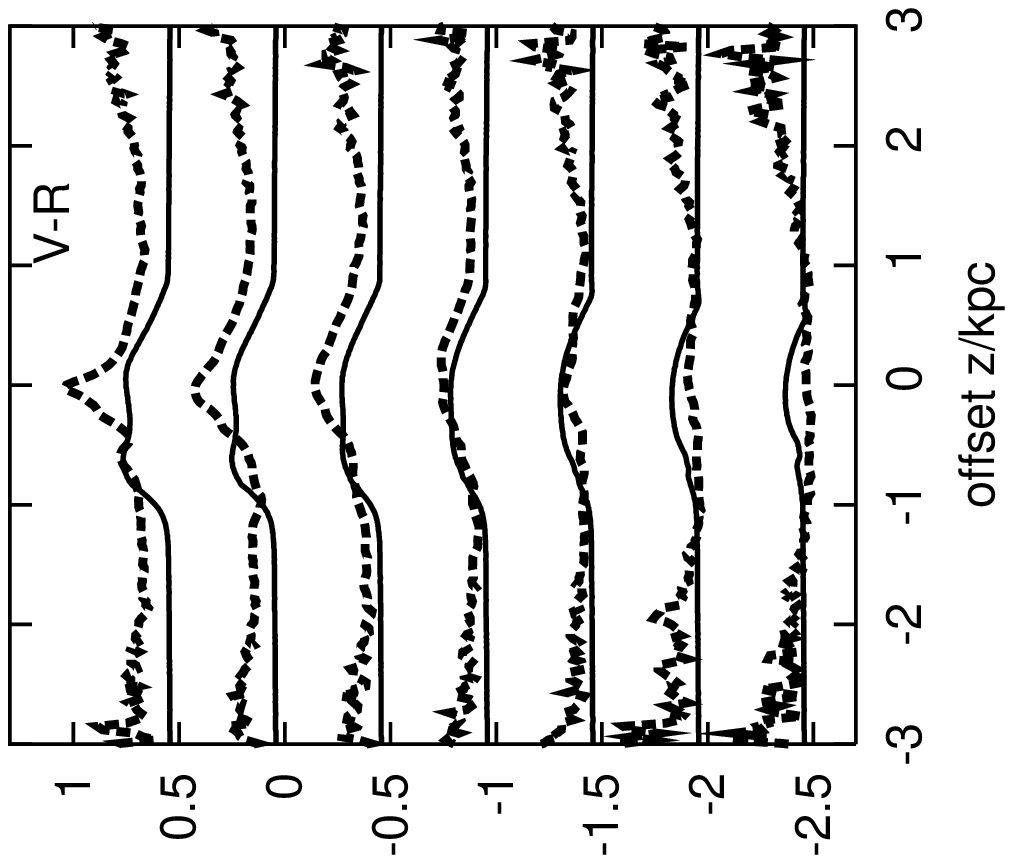}}
  }
\caption[]{
Vertical profiles for NGC 5907 in the U-B, B-V, V-R, and R-I colour
indices from upper left to lower right. The seven profiles in each plot are
the same vertical cuts as in Fig. \ref{figlum}. 
Dashed lines represent the data, while solid lines give
the model. From top to bottom, the magnitude scale is shifted 
by 0, 1, 2, 3, 4, 5 and 6 magnitudes to the blue, respectively.
}
\label{figcol}
\end{figure*}
\subsection{Comparison with the data \label{comparison}}

In Fig. \ref{figlum} the observed and model 
vertical surface-brightness profiles in all five bands are
shown. The five profiles in each plot are
vertical cuts parallel to the minor axis at distances along the major axes 
of 0, 2.4, 4.8, 7.2, 9.6, 12.0, and 14,4\,kpc from top to
bottom, respectively. The general features in the disc-dominated parts are
reproduced reasonably well by the model in all bands. In the inner two profiles
the additional bulge light can be observed and far above the galactic plane an
additional more extended component appears as discussed in 
Lequeux et al. (\cite{leq96}). Only the U-band profiles show a systematic
deviation in the sense that they seem to be radially more extended. The
extinction features near the midplane are of a similar strength as observed but
due to the strong clumping not reproduced in detail. The
large optical depth totally obscures the strong excess stellar light in 
the intrinsic profiles (see Fig. \ref{figlumz}).

The corresponding colour index profiles (Fig. \ref{figcol}) are relatively flat as
expected even at high optical depth due to the mixture of stellar emissivity and
dust extinction along the line of sight (see Just et al. \cite{jus96}). The
systematic shape variation of the reddening features in the 
colour indices U-B, B-V, V-R, and R-I running from asymmetric with
blue dip in the blue colours to more symmetric and redder in the red colours 
is also reproduced.

\subsection{Global properties \label{glob}}

The disc parameters of the final model are collected in Table~\ref{tabmod}.
 The inclination $i=87.5\degr $ and the cutoff radius $R_\mathrm{max}=19.7$\,kpc are very
 similar to the model of Xilouris et al. (\cite{xil99}). 
 The radial and vertical scale length  $R_\mathrm{s}=5.5$\,kpc
and $z_\mathrm{s}=520$\,pc are somewhat larger, because we concentrate our fit
on the height regime $|z|<1.5$\,kpc, where the  Xilouris et al. fit is not
satisfactory.

For the total stellar mass (luminous plus remnants) we obtain from the central
surface density $\Sigma_\mathrm{s,0}=115\,\msun \mathrm{pc}^{-2}$ a total stellar mass of
\bq
M_\mathrm{s}=1.9\times10^{10}\msun 
\eq
corresponding to an average star formation rate of 
$\langle SFR\rangle=M_\mathrm{s}/g_\mathrm{eff}/t_\mathrm{a}=2.34\,\msun$/yr and a present day
value of $SFR_0=1.64\,\msun$/yr.

The dust component has a radial and vertical scale length of 
$R_\mathrm{d}=7.7$\,kpc
and $z_\mathrm{d}=110$\,pc, respectively. The central extinction coefficient is
$A_\mathrm{V,0}=4.0\,$mag/kpc, which corresponds to a face-on optical depth of
$\tau_\mathrm{V}^f=0.81$ and central gas density of
$\rho_\mathrm{d,0}=7.0\times 10^{-4}\,\msun/pc^3$. 
The total dust mass is
$M_\mathrm{d}=4\pi R_\mathrm{d}^2 z_\mathrm{d}\rho_\mathrm{d,0}=5.7\times 10^7\,\msun $ leading to a
gas-to-dust ratio of $F=M_\mathrm{g}/M_\mathrm{d}=209$.

\begin{table}
\begin{tabular}{lr||lr} \hline
 \multicolumn{2}{c}{stellar disc} & \multicolumn{2}{c}{dust component}
         \\ \hline
$i$ & $87.5\degr $ &  &  \\
$R_\mathrm{max}$\,[kpc] & $19.7$ &  &  \\
$R_\mathrm{s}$\,[kpc] & $5.5$ &   $R_\mathrm{d}$\,[kpc] & $7.7$ \\
$z_\mathrm{s}$\,[pc] & $520$ &   $z_\mathrm{d}$\,[pc] & $110$ \\
$S_0$\,[$\msun$ pc$^{-2}$] & 170 & $A_\mathrm{V,0}$\,[mag/kpc] & 4.0 \\
$g_\mathrm{eff}$ & $0.676$ &  &  \\
$C_\mathrm{z}$ & $1.05$ &  &  \\
\hline
\end{tabular}
\caption{Parameters of the final model for NGC 5907. 
At the left hand side the physical quantities of the stellar disc are given 
and at the right hand side the parameters for the dust component. Inclination
$i$ and cutoff radius $R_\mathrm{max}$ are the same for all components. $S_0$ is the
integrated  star formation at the centre and $g_\mathrm{eff}$ is the conversion factor
to the stellar surface density $\Sigma_\mathrm{s,0}=g_\mathrm{eff}S_0=115\,\msun pc^{-2}$.
$C_\mathrm{z}$ is the shape correction factor of the stellar density profile.} 
\label{tabmod}
\end{table}

\subsection{The scaling radius $R_0$ \label{scaling}}

The input parameters of gas and
halo component for the intrinsic disc structure are determined 
at the scaling radius $R_0=10$\,kpc (see Table \ref{tabinp}).
For the normalisation we need the properties of the stellar component from the
final model. The stellar 
surface density is $\Sigma_\mathrm{s}=18.7\,\msun \mathrm{pc}^{-2}$ (Eq. \ref{eqrad}) 
and the maximum velocity dispersion is $\sigma_\mathrm{e}=32$\,km/s (Eq. \ref{eqze}).

For the gas component we adopt the same scale lengths as for the dust, 
i.e. $R_\mathrm{g}=R_\mathrm{d}=7.7$\,kpc and 
$z_\mathrm{g}=z_\mathrm{d}=110$\,pc, which
for the radial distribution is consistent with 
the larger extension of the HI distribution. The vertical scale height 
corresponds to the maximum age bin of $t_\mathrm{g}=3$\,Gyr. This can
be converted to a maximum 
 velocity dispersion of $\sigma_\mathrm{g,m}=0.325\sigma_\mathrm{e}=10.4$\,km/s
and an rms value of $\sigma_\mathrm{g}=0.16\sigma_\mathrm{e}=5.1$\,km/s for
the gas.

For the DM halo the surface density and the velocity dispersion taken 
 from the mass model of Sackett et al. (\cite{sac94}) 
 $\Sigma_\mathrm{h}=48.5\,\msun \mathrm{pc}^{-2}$ 
 up to a maximum height of $z_\mathrm{max}=5.1$\,kpc and
 $\sigma_\mathrm{h}=110$\,km/s (see Sect.~\ref{discparam}) are used.
Since we deduced the halo properties from the rotation curve, we checked
the reconstruction of the rotation curve with our final
model. A mass model leading to an equally good fit of the rotation curve in
Sackett et al. can be constructed by a four component galaxy including a
Hernquist bulge, exponential discs for the stellar and the gas component, 
and a pseudo-isothermal dark matter halo. 
The bulge with a mass of $3\times 10^{10}\,\msun$ and a core radius of 1\,kpc 
reproduces the steep rise of the CO rotation curve presented in Sofue et al.
(\cite{sof96}). We adopt a flattening of 0.5 as determined in Barneby \&
Thronson (\cite{bar94}). The exponential discs are determined by the scalelengths of
5.5\,kpc and 7.7\,kpc and the
total masses of $1.9\times 10^{10}\,\msun$ and $1.2\times 10^{10}\,\msun$
(corresponding to $2.18\times 10^{10}\,\msun$ and $1.65\times 10^{10}\,\msun$
when neglecting the cutoff) for stars and gas, respectively. The dark matter
halo has a local density of $5.5\times 10^{-3}\,\msun/\mathrm{pc}^3$ at
$R_0=10$\,kpc with a core
radius of 1\,kpc. The local density is slightly higher than that used in our
model in order to correct for the effect of the core. 
The rotation curve of NGC~5907 presented in Sofue et al.
(\cite{sof96}) is significantly different and also shows the uncertainties
 in the observational data itself. To reproduce the prominent maximum around
 a radius of 12\,kpc, one can use a NFW halo with scale radius of 6\,kpc and
 local density of $4.78\times 10^{-3}\,\msun/\mathrm{pc}^3$ (as in our model)
 combined with a smaller bulge with core radius 0.5\,kpc and mass of
 $1.2 \times 10^{10}\,\msun$. There is a large freedom
in the parameter choice to reproduce the rotation curves, especially if we allow
the disk mass to vary. Instead of using a NFW profile the maximum in the
rotation curve of Sofue et al. can also be reached by increasing the disc mass 
by a factor of 2--3.
 
The total surface density at $R_0$ is  
$\Sigma_\mathrm{tot}=76.0\,\msun \mathrm{pc}^{-2}$,
which is used to determine the ratios 
$Q_\mathrm{s}=\Sigma_\mathrm{s}/\Sigma_\mathrm{tot}$ etc. 
The halo velocity dispersion is normalised by 
$s_\mathrm{h}=\sigma_\mathrm{h}/\sigma_\mathrm{e}=3.5$.
All input parameters of the final model for NGC 5907 are collected in 
Table~\ref{tabinp}.

\begin{table}
\begin{tabular}{lr||lr} \hline
 \multicolumn{2}{c}{at $R_0=10\,kpc$} & \multicolumn{2}{c}{scaling parameter}
         \\ \hline
$z_\mathrm{s}$[pc] & 520 & $\sigma_\mathrm{e}$[km/s] & 32 \\
$\Sigma_\mathrm{s}[\msun \mathrm{pc}^{-2}]$ & 18.7 & $Q_\mathrm{s}$ & 0.25 \\
$\Sigma_\mathrm{g}$ & 8.8 & $Q_\mathrm{g}$ & 0.11 \\
$\Sigma_\mathrm{h}$ & 48.5 & $Q_\mathrm{h}$ & 0.64 \\
$\sigma_\mathrm{h}$[km/s] & 110 & $s_\mathrm{h}$ & 3.5 \\
$z_\mathrm{g}$[pc] & 110 & $t_\mathrm{g}$[Gyr] & 3.0 
\\ \hline
\end{tabular}
\caption{Input parameters for the final model of NGC 5907. At the left hand side
the physical quantities at $R_0=10$\,kpc are given and at the
right hand side the corresponding normalised parameters of the normalised
profile.
The corresponding description is given in Sect. \ref{scaling}.} 
\label{tabinp}
\end{table}

\subsection{Intrinsic disc properties \label{intrinsic}}

\begin{figure}
\centerline{
  \resizebox{0.98\hsize}{!}{\includegraphics[angle=270]{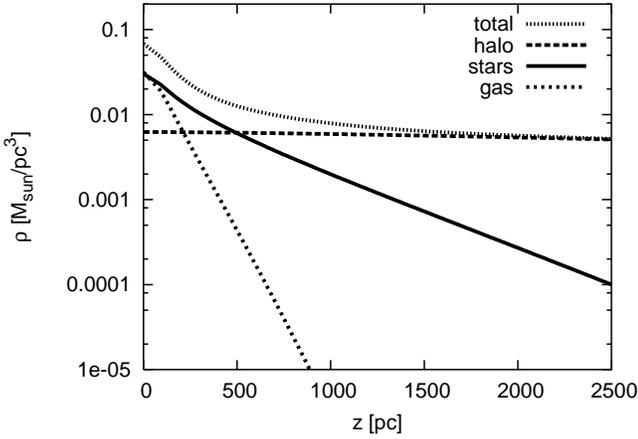}}
  }
\caption[]{
Vertical density profiles of the final of NGC 5907 model normalised to the values 
at the scaling radius $R_0=10$\,kpc. The gas density (short dashed line)
slightly dominates towards the midplane. The stellar density (full line) has an
exponential scale height of $z_\mathrm{s}=520$\,pc with a density enhancement to the
midplane. The halo density (long dashed line) shows only 20\% variation below
$2.5$\,kpc due to the adiabatic contraction.
}
\label{figdens}
\end{figure}

The self-consistent vertical profiles of the final model are shown in Fig.
\ref{figdens}. The stellar density profile is approximately exponential with an
excess below $|z|\approx 500$\,pc. At the midplane the stellar density compared
to the extrapolated exponential profile shows an excess of a factor of 2.1. 
The corresponding
excess in surface density is 48\%. The gas density profile matches the
 scale height of the dust component with 
$z_\mathrm{g}=z_\mathrm{d}=110$\,pc. The central gas density is slightly (5\%) larger than the central
stellar density. 
The DM halo profile is very flat and decreases by 
44\% from the midplane to $z_\mathrm{max}=5.1$\,kpc. 
The halo density
exceeds the disc density (stars plus gas) for $|z|>510$\,pc, but the cumulative
mass, which determines the local gravitational force, dominates only at
$|z|>2.7$\,kpc. 

\begin{table}
\begin{tabular}{clccccc} \hline
 &unit&U&B&V&R&I         \\ \hline
 $\epsilon_0$& mag/pc$^3$  & 6.05&6.17  &5.81   &5.50   &5.10\\
 $(\epsilon/\epsilon_\mathrm{exp})_0$  &-&
         23.3   &15.4   &9.46   &7.59   &6.31\\
$(M/L)_0$ & $(M/L)_\mathrm{\sun}$ 
        &0.27&  0.35    &0.45   &0.56   &0.53  \\
\hline
 $L_0$& mag/pc$^2$  & 0.15      &0.09   &-0.50  &-0.93& -1.43\\
  & mag/$\arcsec^2$  & 22.73     &22.66  &21.05& 20.64&  20.14\\
\hline
 $m_\mathrm{tot}$& mag & 9.65        &9.59   &9.0    &8.57   &8.07 \\
$\Delta m_\mathrm{te}$  & mag & -1.78        &-1.51  &-1.21  &-1.09  &-0.99 \\
$(M/L)_\mathrm{tot}$ & $(M/L)_\mathrm{\sun}$ 
&0.80&0.88&     0.91&1.00&      0.87  \\
$(M/L)_\mathrm{exp}$ & $(M/L)_\mathrm{\sun}$ 
&2.78   &2.39&  1.88    &1.85&  1.46  \\
\hline
\end{tabular}
\caption{Intrinsic properties of the final model of NGC 5907 in the different
bands. The absolute quantities (central emissivity, surface brightness and
total luminosity) depend on the central stellar surface density
and the radial and vertical scale length as given in Table \ref{tabmod}.
The first row gives the central emissivity $\epsilon_0$ 
and the second row shows the enhancement factor 
$(\epsilon/\epsilon_\mathrm{exp})_0$
with respect to the exponential extrapolation form 
$z\approx 3z_\mathrm{s}$ to the midplane. The mass-to-light ratios $(M/L)_0$
at the midplane are very low due to the high fraction young stars.
The next two rows give the central face-on
surface brightness $L(R=0)$ in physical and observational units. $m_\mathrm{tot}$ is
the total apparent luminosity with the distance module $\Delta m=30.2$\,mag 
and $\Delta m_\mathrm{te}=m_\mathrm{tot}-m_\mathrm{exp}$ quantifies the additional
stellar light compared to pure exponential vertical profiles.
The average stellar mass-to-light ratios 
$(M/L)_\mathrm{tot}$ of the model are the same for each vertical
profile and the whole disc. For the exponential extrapolation
$(M/L)_\mathrm{exp}$ is constant all over the disc and equals the 
$M/L$ of the model at large heights $z$.
}\label{tablum}
\end{table}

\begin{figure*}[t]
\centerline{
  \resizebox{0.49\hsize}{!}{\includegraphics[angle=270]{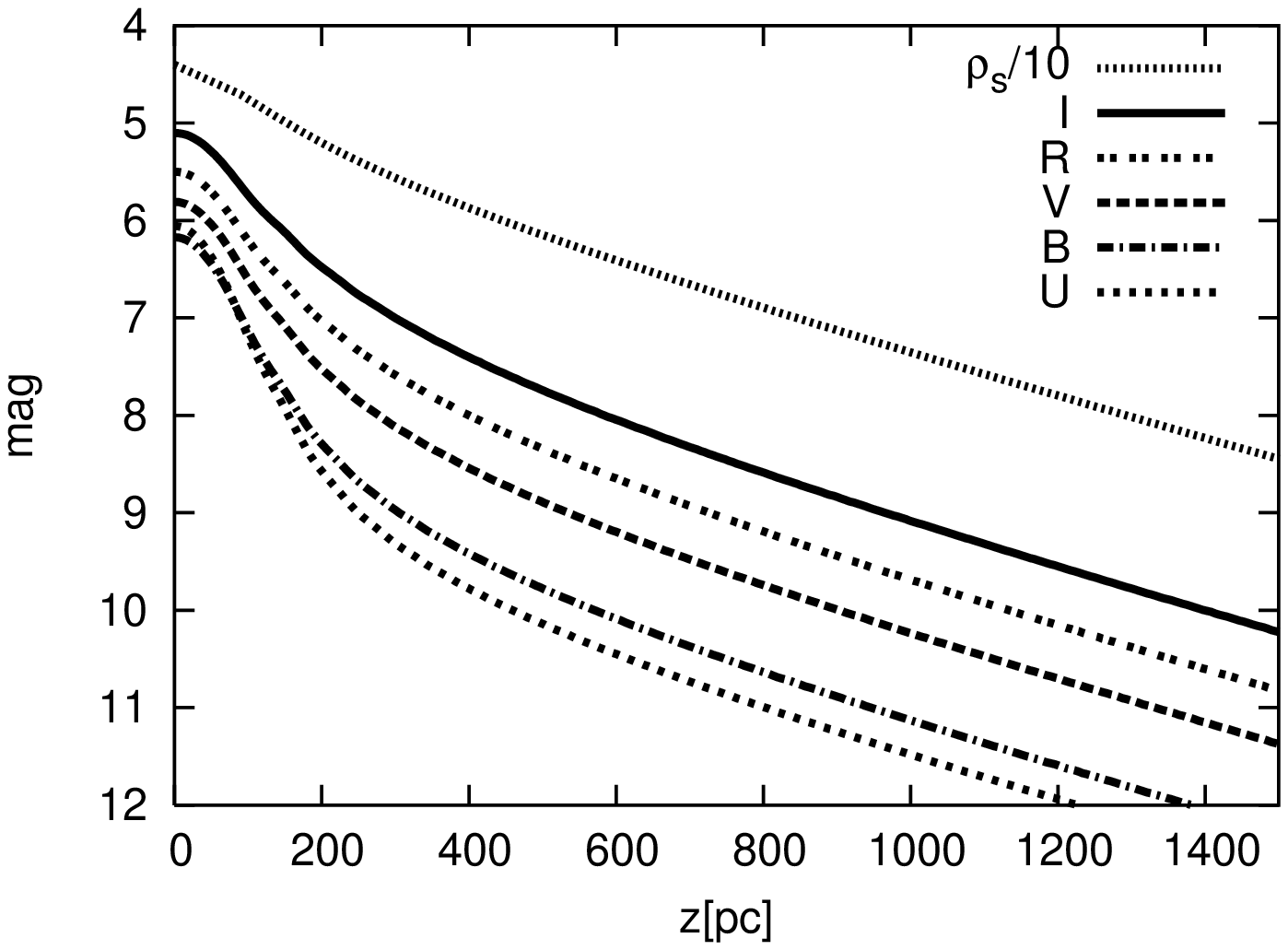}}
  \resizebox{0.49\hsize}{!}{\includegraphics[angle=270]{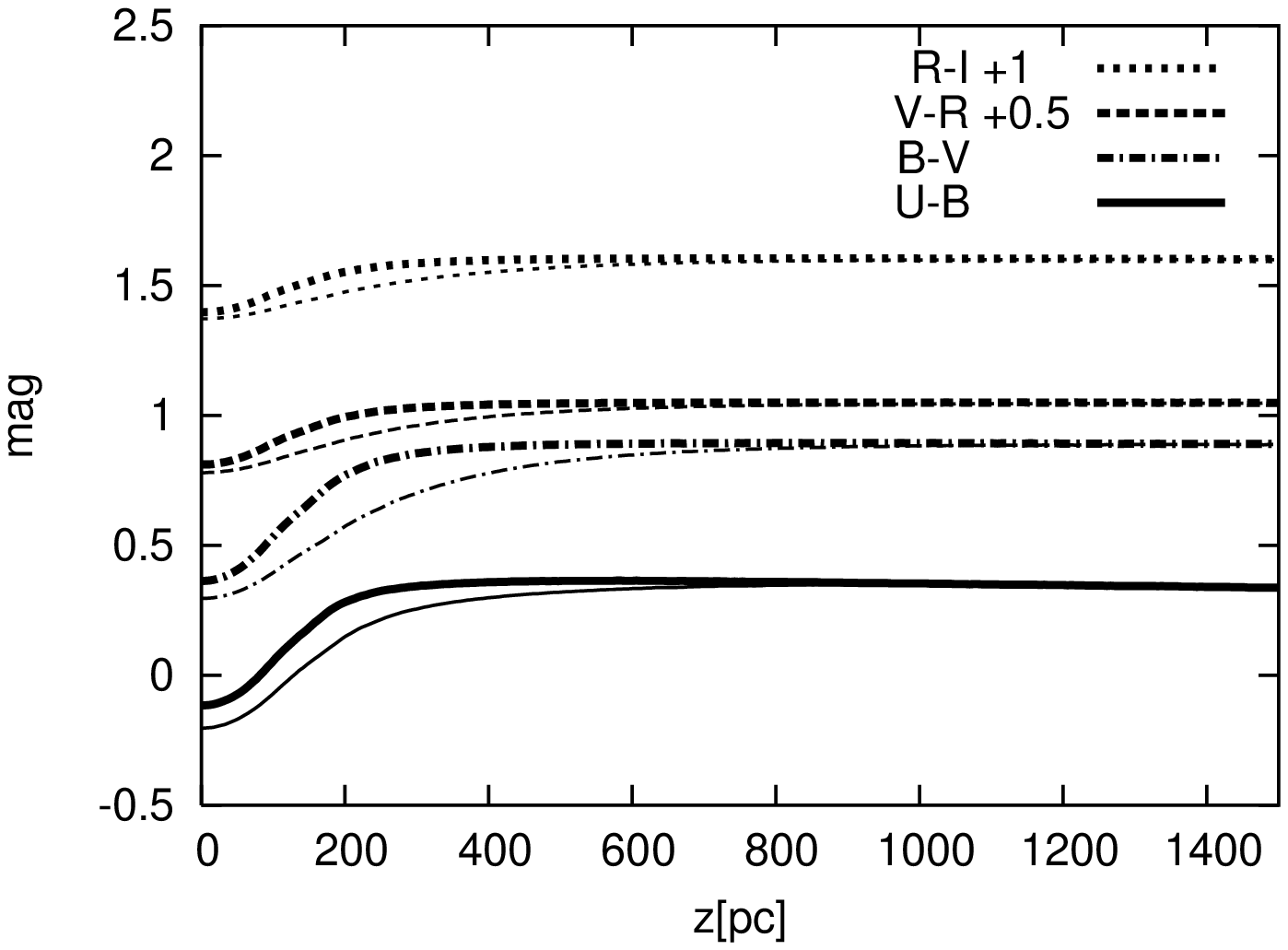}}
  }
\centerline{
  \resizebox{0.49\hsize}{!}{\includegraphics[angle=270]{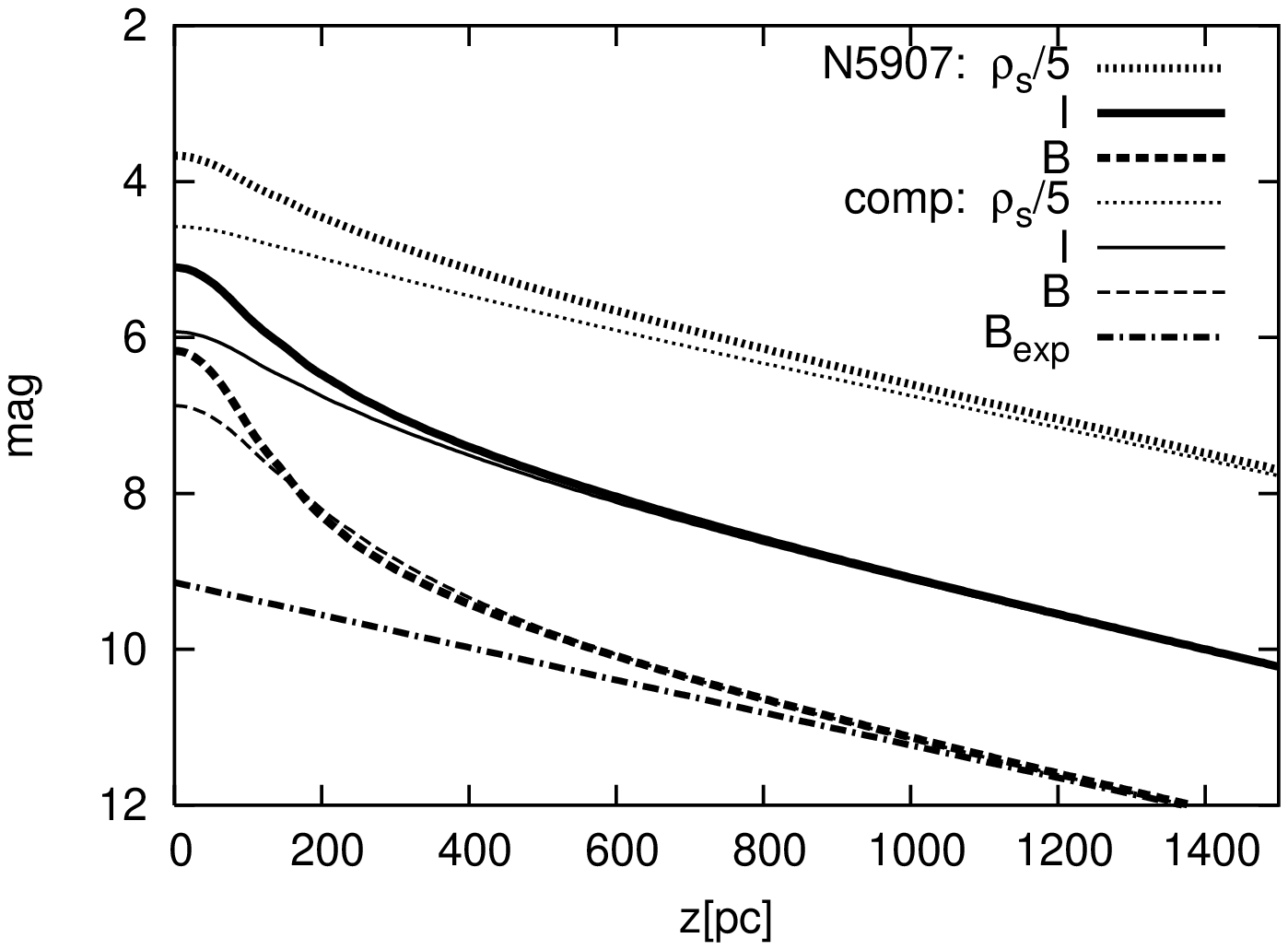}}
  \resizebox{0.49\hsize}{!}{\includegraphics[angle=270]{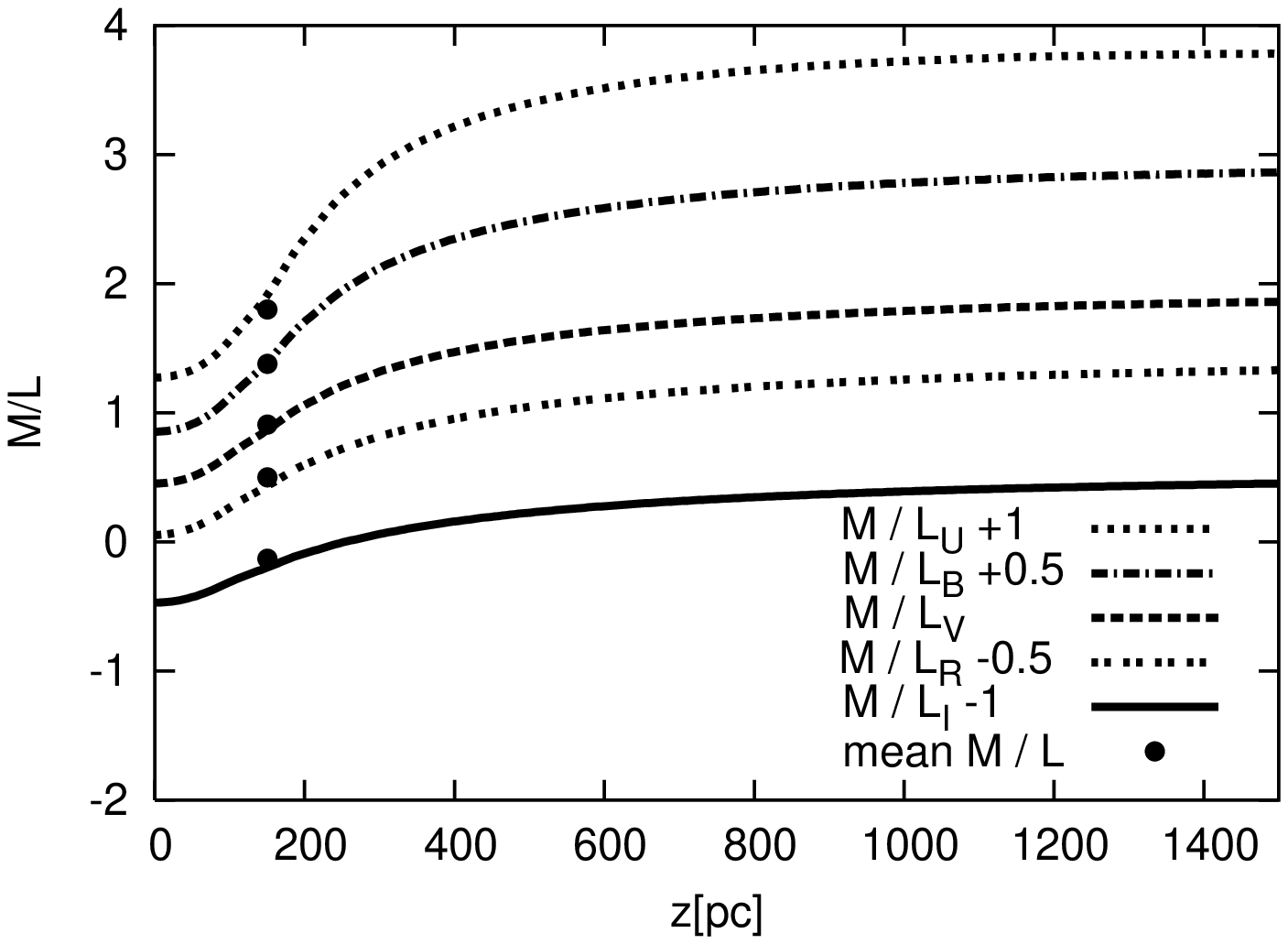}}
  }
\caption[]{
{\bf Upper left panel:} The emissivity of the central profile of NGC
5907 in $mag\,pc^{-3}$. For comparison the stellar density profile is also
plotted in units of $\msun \,\mathrm{pc}^{-3}/10$ (also in mag). 
{\bf Lower left panel:} The B and
I-band emissivity for comparison with a model using the heating function of the
solar neighbourhood and a constant star formation rate. For the B-band profile
the exponential profile fit (dot-dashed line) is shown. The
corresponding density profiles are normalised to $\msun \,\mathrm{pc}^{-3}/5$.
{\bf Upper right panel:} Intrinsic colour profiles at the centre of NGC
5907 (thick lines). The profiles are shifted to avoid overlap. 
The profiles of the comparison model (thin lines) with constant $SFR$ 
show more extended
colour gradients due to the faster dynamical heating (see Fig. \ref{figsfr}).
{\bf Lower right panel:} The mass-to-light ratios in all bands, also shifted to
avoid overlap. The dots denote the average values for the total surface density
(see Table \ref{tablum}). 
}
\label{figlumz}
\end{figure*}

The left panels in Fig. \ref{figlumz} show the intrinsic luminosity profiles in 
all bands.
For comparison the density profile is also plotted to emphasise the enhanced
stellar light of the young stars near the midplane. As the lower panel shows for
the B-band, the profiles deviate from the exponential shape already at heights
below $|z|=1$\,kpc. In the right panels the corresponding colour profiles and
mass-to-light ratios are shown. The intrinsic colours above
$|z|\approx 200$\,pc show very shallow gradients for this model. 
The colour gradient due to the
 blue young population can be seen only in the innermost part. The mass-to-light
 ratio is more sensitive to the age-distribution of the stars and varies
 much more strongly than the colours. It varies by a
 factor of 10 in U-band and even in I-band by a factor of 3. The
 $M/L$-gradient is significantly
 larger than zero up to $|z|\approx500$\,pc at least in the blue bands. 
 The average value of $M/L$
 corresponds to the local value near a height of $|z|\approx 150$\,pc (dots in
 the lower right panel of Fig. \ref{figlumz}).

The luminosity weighted mean metalicity
of the final model varies from  $[Fe/H]=0.06$ at the midplane to 
$[Fe/H]=-0.59$ at high latitude $|z|$, which is only slightly less than the 
variation in the chemical enrichment of the subpopulations.
 
In order to show that the evolutionary history cannot be decomposed arbitrarily
into a dynamical heating function and an appropriate star formation history, we
calculated comparison models, where we have fixed the heating function by that of 
the solar neighbourhood (Fig. \ref{figsfr} lower panel, thin line)
and changed the $SFR$ in order to get a higher fraction of
young stars. But even with a constant $SFR$ the strong heating of the young
subpopulations spread the blue stars to a much larger height yielding more
extended colour gradients and a smaller stellar light excess near the midplane
(thin lines in Fig. \ref{figlumz}).

\section{Discussion \label{discuss}}

\subsection{Disc evolution \label{evol}}

Our final model combines a star formation history with a moderate maximum at an
age of $t=7.2$\,Gyr with a slow dynamical heating for the young stellar
population compared to the solar neighbourhood (see Fig. \ref{figsfr}). If we
use as in the comparison model the heating function of the solar neighbourhood
(constant increase of kinetic energy of the subpopulations with age), a higher
present day $SFR$ is necessary to get a comparable luminosity in the blue bands
near the midplane. Even a constant $SFR$ 
(which is often claimed even for the Milky
Way) does not reach the intrinsic luminosity of our final model, but
leads to additional blue light at intermediate heights $|z|$, where dust
reddening cannot compensate. We do not claim that the chosen $SFR$ is
unique, because the parameter space is too large and a model on a much higher
level of comparison including radial variation of the intrinsic profiles and
clumping near the midplane would be necessary. Our result shows that a
continously decreasing $SFR$ in the last few Gyr can account for the observed
high intrinsic luminosity at the midplane necessary for the strong NIR-submm
brightness. A recent star burst is not needed but also not excluded by our model
at this level of modelling using a simple radial disc structure.

\subsection{Mass-to-light ratio \label{ml}}

The stellar mass-to-light ratio $M/L$ strongly depends on the $IMF$ at the low
mass end without changing the luminosities much. In our model with the 
Scalo-IMF we get a relatively low value $M/L_\mathrm{V}=0.91$, which fits well to the
mass models of Sackett et al. (\cite{sac94}). Since the disc of an Sc galaxy is
younger than an Sbc galaxy like the Milky Way, this value also fits well
compared to the values in the solar neighbourhood. 
A rough estimate of the excess in blue light of the young stellar population in
our model compares well to the additional light needed in
Misiriotis et al. (\cite{mis01}) to heat the dust. 
Their corresponding mass-to-light ratio and recent $SFR$ are slightly different,
because they used a different stellar population model (Salpeter $IMF$ and 
no mass loss due to stellar evolution). 

The low mass-to-light ratio of the stellar disc is a clear indication 
of a low disc mass model
 as discussed also in Sackett et al. (\cite{sac94}). Our underlying mass model
 is consistent with the observed rotation curve of Sackett et al. with an
Pseudo-isothermal DM halo and 
for the rotation curve of Sofue et al. (\cite{sof96}) a   NFW halo fits better. 
But since the
mass model is fully self-consistent at the scale radius $R_0=10\,$kpc only, 
we cannot distinguish between isothermal or NWF profiles for the
dark matter halo.

\subsection{Dust component \label{dustcomp}}

For the dust component we find a similar scale height as Misiriotis et al.
(\cite{mis01}) and a similar total dust mass, which is responsible for the FIR
brightness. A significant difference to the models of Xilouris et al.
(\cite{xil99}) and Misiriotis et al. is the larger dust scale length of
$R_\mathrm{d}=7.7$\,kpc. The effect of a larger $R_\mathrm{d}$ is, apart from the stronger
extinction feature at large distance form the minor axis, 
a shift of the extinction dip to
higher vertical offset at all vertical cuts. This effect cannot be substituted
by a larger inclination $i$, because this would smear out the extinction feature
to larger heights above the major axis. A side effect of the larger radial
scale length is the reduced face-on optical depth at the centre 
($\tau_\mathrm{V}^f=0.81$ instead of 1.5). Xilouris et al. also found that the dust
scale length is larger than the stellar scale length in their best fit. The
large radial scale length of the dust also fits to 
the more extended distribution of HI and of the FIR brightness.

\section{Conclusions \label{conclusion}}

We have observed deep photometry of NGC 5907 in the U,
B, V, R, and I-band in order to construct a physical disc model based on a
multi-colour analysis. 
We have constructed a self-consistent evolutionary disc model for NGC 5907,
which fits the general characteristics of the surface brightness distribution in
all bands. The stellar disc is built up by a sequence of stellar
subpopulations distributed according to a smooth $SFR$ and dynamical heating
function $\sigma(t)$. The emissivity is calculated with stellar population
synthesis and the appearance of the galaxy is calculated by radiative transfer 
including an exponential disc component of dust. 

This is the first physical model of the disc
of NGC 5907 that gives the correct luminosities and colours of the outer
exponential parts of the disc and also the large amount of dust observed in
FIR/submm. The reason for the corresponding excess in stellar light near the
midplane of the disc is the concentrated luminous young subpopulation 
produced by the continuous $SFR$. 
Our model is a significant improvement of the disc model of
Misiriotis et al. (\cite{mis01}), who combined the phenomenological exponential
model of Xilouris et al. (\cite{xil99}) with an additional thin disc 
of uniformly mixed
young stars and dust in order to produce the excess in FIR/submm luminosity.

This evolutionary disc model naturally explains three different aspects known
from the literature:
1) The exponential vertical structure well above the midplane as modelled
phenomenologically by Xilouris et al.(\cite{xil99}) 
is the result of the older subpopulations in dynamical equilibrium in the total
disc potential.
2) The large amount of dust observed in the FIR/submm 
(modelled by Misiriotis et al. \cite{mis01}) is necessary to hide the bright
young subpopulations confined to the midplane. The young stars are the smooth
end of the $SFR$ without a recent star burst.
3) The additional light necessary to heat the dust is automatically delivered by
the high mass-to-light ratio near the midplane.

\section*{Acknowledgements}
 
We thank the Deutsche Forschungsgemeinschaft for supporting
this project by the SFB~328 and SFB~439 at the University of
Heidelberg. We are grateful to the staff of the Calar Alto Observatory
for the support during the observations.



\end{document}